\documentclass[]{elsarticle}
\usepackage[dvipsnames]{xcolor}
\usepackage[utf8]{inputenc}

\makeatletter
\def\ps@pprintTitle{%
 \let\@oddhead\@empty
 \let\@evenhead\@empty
 \def\@oddfoot{}%
 \let\@evenfoot\@oddfoot}
\makeatother

\usepackage{moreverb}

\usepackage[colorlinks,bookmarksopen,bookmarksnumbered,citecolor=red,urlcolor=red]{hyperref}
\usepackage{geometry}
\renewcommand{\vec}[1] {\ensuremath{\mathbf #1}}
\newcommand{\gvec}[1] {\ensuremath{\boldsymbol #1}}

\newcommand{\n} {\ensuremath{\vec{n}}}
\newcommand{\I} {\ensuremath{\vec{I}}}

\newcommand{\V} {\ensuremath{\vec{V}}}

\newcommand{\dprod}{{\scriptscriptstyle\bullet}}

\newcommand{\tdprod}%
     {\,{\scriptscriptstyle \stackrel{3}{\bullet}}\,}
\renewcommand{\max} {\ensuremath{\operatorname{max}}}

\newcommand{\dV} {\ensuremath{\,dV}}

\newcommand{\dS} {\ensuremath{\,dS}}

\newcommand{\dL} {\ensuremath{\,dL}}

\newcommand{\grad}{\ensuremath{\nabla}}

\newcommand{\sGrad} {\ensuremath{\nabla\!_s}}

\newcommand{\ddt}[1] {\ensuremath{\frac{\partial #1}{\partial t }}}


\makeatletter
\DeclareFontFamily{OMX}{MnSymbolE}{}
\DeclareSymbolFont{myLargesymbols}  {OMX}{MnSymbolE}{m}{n}
\SetSymbolFont{myLargesymbols}{bold}{OMX}{MnSymbolE}{b}{n}
\DeclareFontShape{OMX}{MnSymbolE}{m}{n}{
    <-6>  MnSymbolE5
   <6-7>  MnSymbolE6
   <7-8>  MnSymbolE7
   <8-9>  MnSymbolE8
   <9-10> MnSymbolE9
  <10-12> MnSymbolE10
  <12->   MnSymbolE12}{}
\DeclareFontShape{OMX}{MnSymbolE}{b}{n}{
    <-6>  MnSymbolE-Bold5
   <6-7>  MnSymbolE-Bold6
   <7-8>  MnSymbolE-Bold7
   <8-9>  MnSymbolE-Bold8
   <9-10> MnSymbolE-Bold9
  <10-12> MnSymbolE-Bold10
  <12->   MnSymbolE-Bold12}{}
  
\DeclareMathSymbol{\downbrace}    {\mathord}{myLargesymbols}{'251}
\DeclareMathSymbol{\downbraceg}   {\mathord}{myLargesymbols}{'252}
\DeclareMathSymbol{\downbracegg}  {\mathord}{myLargesymbols}{'253}
\DeclareMathSymbol{\downbraceggg} {\mathord}{myLargesymbols}{'254}
\DeclareMathSymbol{\downbracegggg}{\mathord}{myLargesymbols}{'255}
\DeclareMathSymbol{\upbrace}      {\mathord}{myLargesymbols}{'256}
\DeclareMathSymbol{\upbraceg}     {\mathord}{myLargesymbols}{'257}
\DeclareMathSymbol{\upbracegg}    {\mathord}{myLargesymbols}{'260}
\DeclareMathSymbol{\upbraceggg}   {\mathord}{myLargesymbols}{'261}
\DeclareMathSymbol{\upbracegggg}  {\mathord}{myLargesymbols}{'262}
\DeclareMathSymbol{\braceld}      {\mathord}{myLargesymbols}{'263}
\DeclareMathSymbol{\bracelu}      {\mathord}{myLargesymbols}{'264}
\DeclareMathSymbol{\bracerd}      {\mathord}{myLargesymbols}{'265}
\DeclareMathSymbol{\braceru}      {\mathord}{myLargesymbols}{'266}
\DeclareMathSymbol{\bracemd}      {\mathord}{myLargesymbols}{'267}
\DeclareMathSymbol{\bracemu}      {\mathord}{myLargesymbols}{'270}
\DeclareMathSymbol{\bracemid}     {\mathord}{myLargesymbols}{'271}

\def\horiz@expandable#1#2#3#4#5#6#7#8{%
  \@mathmeasure\z@#7{#8}%
  \@tempdima=\wd\z@
  \@mathmeasure\z@#7{#1}%
  \ifdim\noexpand\wd\z@>\@tempdima
    $\m@th#7#1$%
  \else
    \@mathmeasure\z@#7{#2}%
    \ifdim\noexpand\wd\z@>\@tempdima
      $\m@th#7#2$%
    \else
      \@mathmeasure\z@#7{#3}%
      \ifdim\noexpand\wd\z@>\@tempdima
        $\m@th#7#3$%
      \else
        \@mathmeasure\z@#7{#4}%
        \ifdim\noexpand\wd\z@>\@tempdima
          $\m@th#7#4$%
        \else
          \@mathmeasure\z@#7{#5}%
          \ifdim\noexpand\wd\z@>\@tempdima
            $\m@th#7#5$%
          \else
           #6#7%
          \fi
        \fi
      \fi
    \fi
  \fi}

\def\overbrace@expandable#1#2#3{\vbox{\m@th\ialign{##\crcr
  #1#2{#3}\crcr\noalign{\kern2\p@\nointerlineskip}%
  $\m@th\hfil#2#3\hfil$\crcr}}}
\def\overbrace@#1#2#3{\vbox{\m@th\ialign{##\crcr
  #1#2\crcr\noalign{\kern2\p@\nointerlineskip}%
  $\m@th\hfil#2#3\hfil$\crcr}}}
  
\def\underbrace@expandable#1#2#3{\vtop{\m@th\ialign{##\crcr
  $\m@th\hfil#2#3\hfil$\crcr
  \noalign{\kern2\p@\nointerlineskip}%
  #1#2{#3}\crcr}}}
\def\underbrace@#1#2#3{\vtop{\m@th\ialign{##\crcr
  $\m@th\hfil#2#3\hfil$\crcr
  \noalign{\kern2\p@\nointerlineskip}%
  #1#2\crcr}}}

\def\bracefill@#1#2#3#4#5{$\m@th#5#1\leaders\hbox{$#4$}\hfill#2\leaders\hbox{$#4$}\hfill#3$}

\def\downbracefill@{\bracefill@\braceld\bracemd\bracerd\bracemid}
\DeclareRobustCommand{\downbracefill}{\downbracefill@\textstyle}

\def\upbracefill@{\bracefill@\bracelu\bracemu\braceru\bracemid}
\DeclareRobustCommand{\upbracefill}{\upbracefill@\textstyle}

\def\upbrace@expandable{%
  \horiz@expandable
    \upbrace
    \upbraceg
    \upbracegg
    \upbraceggg
    \upbracegggg
    \upbracefill@}
\def\downbrace@expandable{%
  \horiz@expandable
    \downbrace
    \downbraceg
    \downbracegg
    \downbraceggg
    \downbracegggg
    \downbracefill@}
    
\DeclareRobustCommand{\overbrace}[1]{\mathop{\mathpalette{\overbrace@expandable\downbrace@expandable}{#1}}\limits}
\DeclareRobustCommand{\underbrace}[1]{\mathop{\mathpalette{\underbrace@expandable\upbrace@expandable}{#1}}\limits}
\makeatother
\font\bigtenrm=cmr12 scaled 1200
\newcommand{\eexp}[1]{{\hbox{$\textfont1=\bigtenrm e$}}^{\raise3pt\hbox{$#1$}}}

\newcommand{\file}[1]{\textit{#1}}
\newcommand{\parameter}[1]{\lstinline{#1}} % lstinline or verb

\newcommand{\vel}{\mathbf{v}}

\graphicspath{{figures/}}
\usepackage{subcaption}
\usepackage{algorithm}
\usepackage{algpseudocode}
\usepackage[numbers]{natbib}
\usepackage{import} % use by inkscape-produced figures with latex content
\usepackage{siunitx}
\usepackage[textsize=tiny,disable]{todonotes}
\usepackage{amssymb}
\usepackage{amsthm}
\usepackage{mathtools} % added by SR
\usepackage{amsmath}
\usepackage{booktabs}
\usepackage{longtable}
\usepackage{cleveref}
\usepackage{lineno}
\usepackage{txfonts} % added by SR
\usepackage{listings} % addedhttps://www.openfoam.com/ by MS (minted would be nice, too)
\usepackage{lmodern} % added by MS (for textit)

\usepackage{soul}

\colorlet{Reviewer1}{Black}%Cerulean}
\newcommand{\ReviewerOne}[1]{\textcolor{Reviewer1}{#1}}
\colorlet{Reviewer2}{Black}%OliveGreen}
\newcommand{\ReviewerTwo}[1]{\textcolor{Reviewer2}{#1}}

\newcommand{\otherChange}[1]{\textcolor{Black}{#1}}%RawSienna}{#1}}

\let\vec\mathbf

\begin{document}

\begin{frontmatter}

\title{twoPhaseInterTrackFoam: an OpenFOAM module for Arbitrary Lagrangian/Eulerian Interface Tracking with Surfactants and Subgrid-Scale Modeling}

\affiliation[addrTUDa]{organization={Mathematical Modeling and Analysis Institute, Mathematics department, TU Darmstadt}, country={Germany}}
%Peter-Grünberg-Straße 10, 64287 Darmstadt, Germany}

\affiliation[addrUniZ]{organization={Faculty of Mechanical Engineering and Naval Architecture, University of Zagreb}, 
country={Croatia}}
%Ivana Lučića 5, HR-10002 Zagreb, Croatia}

\author[addrTUDa]{Moritz Schwarzmeier}
\ead{schwarzmeier@mma.tu-darmstadt.de}

\author[addrTUDa]{Suraj Raju}
\ead{raju@mma.tu-darmstadt.de}

\author[addrUniZ]{\v{Z}eljko~Tukovi\'c} 
\ead{zeljko.tukovic@fsb.unizg.hr}

\author[addrTUDa]{Mathis~Fricke} 
\ead{fricke@mma.tu-darmstadt.de}

\author[addrTUDa]{Dieter~Bothe} 
\ead{bothe@mma.tu-darmstadt.de}

\author[addrTUDa]{Tomislav~Mari\'{c}\,\corref{corr}} 
\ead{maric@mma.tu-darmstadt.de}
\cortext[corr]{Corresponding author}

\begin{keyword}
%% keywords here, in the form: keyword \sep keyword

%% PACS codes here, in the form: \PACS code \sep code

%% MSC codes here, in the form: \MSC code \sep code
%% or \MSC[2008] code \sep code (2000 is the default)

finite volume \sep interface tracking \sep ALE \sep unstructured mesh \sep subgrid-scale 
\end{keyword}

\pagenumbering{arabic}
\frenchspacing

\begin{abstract}

We provide an implementation of the unstructured Finite-Volume Arbitrary Lagrangian / Eulerian (ALE) Interface-Tracking method for simulating incompressible, immiscible two-phase flows as an OpenFOAM module. In addition to interface-tracking capabilities that include tracking of two fluid phases, an implementation of a Subgrid-Scale (SGS) modeling framework for increased accuracy when simulating sharp boundary layers is enclosed. The SGS modeling framework simplifies embedding subgrid-scale profiles into the unstructured Finite Volume discretization. Our design of the SGS model library significantly simplifies adding new SGS models and applying SGS modeling to Partial Differential Equations (PDEs) in OpenFOAM. 
%
% In addition to the archive published alongside this manuscript \citep{tudatalib}, an actively developed source code repository is available on GitLab\citep{gitlabrepo}. The publication comes with continuous integration (CI) and all the results presented are produced within the CI.

\vspace{3mm}
\textbf{\textcolor{Red}{This is an arXiv preprint. Please refer to and cite the published article\\
M. Schwarzmeier, S. Raju, Ž. Tuković, M. Fricke, D. Bothe, and T. Marić. twoPhaseInterTrackFoam: an OpenFOAM module for Arbitrary Lagrangian/Eulerian Interface Tracking with Surfactants and Subgrid-Scale Modeling. Computer Physics Communications, 308:109460, Mar. 2025. doi:\href{https://doi.org/10.1016/j.cpc.2024.109460}{10.1016/j.cpc.2024.109460}}}

\vspace{5mm}
\noindent{\bf PROGRAM SUMMARY}
  %Delete as appropriate.

\begin{small}
\noindent
{\em Program Title:} twoPhaseInterTrackFoam                                          \\
{\em CPC Library link to program files:} (to be added by Technical Editor) \\
{\em Developer's repository link:} \url{https://gitlab.com/interface-tracking/twophaseintertrackfoamrelease} \\
{\em Code Ocean capsule:} (to be added by Technical Editor)\\
{\em Licensing provisions:} GPLv3 \\
{\em Programming language:} C++\\
%{\em Supplementary material:}                                 \\
  % Fill in if necessary, otherwise leave out.
{\em Nature of problem:}\\
  Two-phase flow problems involving surface-active agents (surfactants), variable surface tension force and very sharp boundary layers.\\
{\em Solution method:}\\
  An OpenFOAM implementation of the Arbitrary Lagrangian / Eulerian Interface Tracking method. \\
% {\em Additional comments including restrictions and unusual features:}\\
%   %Provide any additional comments here  (approx. 50-250 words)
%   TODO
   %\\
%\begin{thebibliography}{0}
%%\bibitem{1}Reference 1         % This list should only contain those items referenced in the                 
%\bibitem{2}Reference 2         % Program Summary section.   
%\bibitem{3}Reference 3         % Type references in text as [1], [2], etc.
                               % This list is different from the bibliography at the end of 
                               % the Long Write-Up.
%\end{thebibliography}
%* Items marked with an asterisk are only required for new versions
%of programs previously published in the CPC Program Library.\\
\end{small}

\end{abstract}

% All CPiP articles must contain the following
% PROGRAM SUMMARY.

%\maketitle

\end{frontmatter}

%\linenumbers
\section{Introduction}
\label{sec:intro}

The unstructured Finite Volume Arbitrary Lagrangian / Eulerian Interface Tracking  method in OpenFOAM (ALE-IT) \citep{Tukovi2012,Tukovi2018} is a highly accurate method that tracks fluid phases as deforming solution domains separated by \ReviewerOne{ fluid interfaces in the form of deforming domain boundaries (cf. \cref{fig:interfaceTracking}).} \ReviewerOne{Modeling fluid interfaces using domain boundaries makes it possible to directly apply jump conditions at fluid interfaces as boundary conditions.} \ReviewerOne{Additionally, it is possible to solve Partial  Differential Equations (PDEs) on the fluid interface, by discretizing surface transport equations using the discretization of the domain boundary.}
Our implementation of the ALE-IT method, which we provide in this manuscript, has been successfully applied to hydrodynamically challenging two-phase flow problems \citep{marschall_validation_2014}, two-phase flows with soluble surfactant \citep{Dieter2014,Dieter2015,pesci_experimental_2017_book} and two-phase flows with interfacial mass transfer \citep{Weber2017,pesci_SGS_risBubb_surfactants_2018}.
\ReviewerTwo{This type of the ALE-IT method has also been extended to handle large mesh deformations by allowing for topological changes on tetrahedral meshes, driven by mesh quality criteria \citep{quan_modeling_2009,menon_parallel_2015}. However, transferring the methodology from \citep{menon_parallel_2015} into our implementation of the ALE-IT method is a complex task, which we will address in our future work.}

\ReviewerOne{Mass transfer across the fluid interface often results in extremely steep concentration gradients, that cannot be resolved by modern numerical simulation methods at affordable computational costs.} The Subgrid-Scale (SGS) modeling approach \cite{Alke_DNS-VoF_2010,bothe_fleckenstein_VoF-SGS_2013,weiner_advanced_2017} employs a \ReviewerOne{local} analytical solution to enhance the \ReviewerOne{computation accuracy for the} passive scalar transport in interfacial boundary layers, saving multiple local refinement levels \ReviewerOne{in the} Finite Volume mesh.
This \ReviewerOne{speeds} up simulations \ReviewerOne{significantly}, \ReviewerOne{in some cases enabling them in the first place}, \ReviewerOne{in other cases saving computational resources}, while delivering results with very high accuracy.
\ReviewerOne{The topic of Subgrid-Scale modeling for transport processes at fluid interfaces is gaining attention and some modifications have been made to include reactive species \citep{grunding_reactive_SGS_2016} or local curvature effects \citep{grosso_thermal_2024}.
While the SGS modeling addressed in this paper is implemented in the ALE-IT method, others have implemented it in the VoF method, e.g. \citep{weiner_advanced_2017,cai_sub-grid_2021}, or in the Front Tracking method \citep{grosso_thermal_2024,claassen_improved_2020}.}

With this publication, we provide an open-source implementation of our ALE-IT method as an OpenFOAM module, \ReviewerOne{which simplifies the development of SGS models for interfacial transport phenomena in OpenFOAM. Researchers that do not use OpenFOAM will be able to analyze our implementation details, which is very helpful for developing these methods.}

\ReviewerOne{The version of the ALE-IT method from} this manuscript is \textit{v1.\otherChange{1}}, available in the GitLab repository \citep{GitlabRepository} and archived as \citep{tudatalib_code}. 
Results reported in this paper are archived as \citep{tudatalib_data}.\todo{update tags, upload data, update links}

% MS: "The advantages of version \textit{v1.0} of our ALE-IT method..."
% I removed the version, the advantages are not bound to it.
\ReviewerOne{The advantages of our ALE-IT method are high simulation accuracy of of two-phase flows involving incompressible fluid phases with strong differences in densities, strong influence of surface tension forces, effects of variable surface tension coefficients, transport of species concentrations on and across the fluid interface, and the Subgid-Scale modeling of mass transfer at fluid interfaces.} 

\ReviewerTwo{This version of the ALE-IT method is limited to moderate deformations of the fluid interface and does not allow for topological changes of the fluid interface, which will be addressed in our future work. Compared to other two-phase flow simulation methods, including those available in OpenFOAM, our ALE-IT method excells in accuracy, under above mentioned limitations.}

The following sections cover the details of the mathematical model, equation discretization, and \otherChange{Subgrid-Scale} modeling implemented in the version \textit{1.\otherChange{1}} of our ALE-IT OpenFOAM module, as well as the results for some showcases.

\section{Mathematical model}
\label{sec:mathModel}

\subsection{Arbitrary Lagrangian/Eulerian (Sharp) Interface (Tracking) Model}
%\label{secSub:??}

\begin{figure}[hptb]
    \centering 
    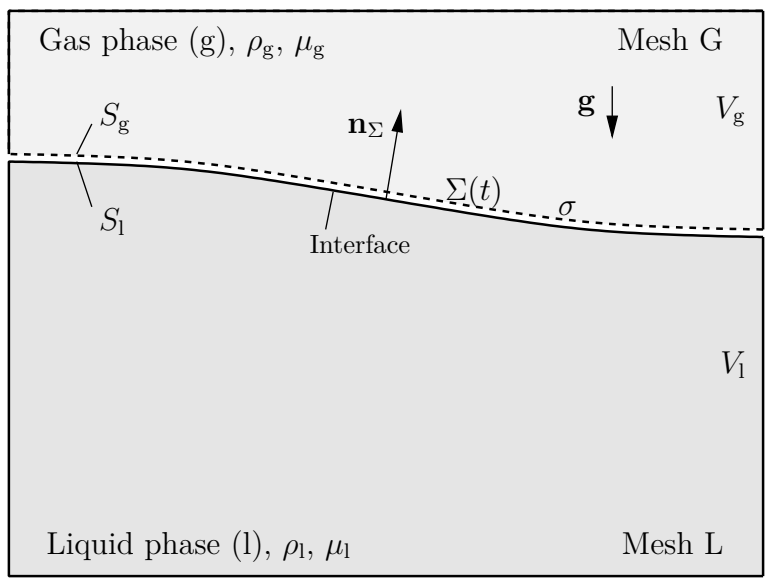 
    \caption{Definition of the solution domain for the ALE-IT method.}
\label{fig:interfaceTracking}
\end{figure}

%A two-phase fluid flow with a sharp interface is simulated using the finite volume (FV/FVM) method and a moving computational mesh, where one defines a separate computational mesh on each phase and these meshes move and deform according to interface motion. 

We model isothermal flow of two immiscible incompressible fluids, schematically shown in \cref{fig:interfaceTracking}, e.g.\ as a liquid and a gas phase separated by a sharp interface $\Sigma(t)$.
Appropriate boundary conditions are then enforced at the fluid interface $\Sigma(t)$, see \citep{Tukovi2012}. We only briefly outline the model, the reader is additionally referred to \citep{slattery_interfacial_2007} for more details.

Isothermal flow of an incompressible Newtonian fluid inside an arbitrary moving control volume~$V$ bounded by a closed surface $S :=\partial V $, i.e.~$S$ is the boundary of the finite volume, moving with the boundary velocity~$\vel_s$ is governed by the mass and linear momentum conservation laws:
\begin{equation}
  \oint\limits_{S}\rho  \vel \cdot \n \,\dS = 0,
  \label{eq:massCLIncompressible}
\end{equation}
\begin{equation}
  %\begin{split}
       \ddt{}\int\limits_{V}\rho\vel\,\dV
      + \oint\limits_{S}\rho((\vel - \vel_s)\otimes\vel) \cdot \n \,\dS %\\
     = \oint\limits_{S}\gvec{\tau}\cdot \n \,\dS
      - \int\limits_{V}\grad p\,\dV,
  %\end{split}
      \label{eq:momentumCLIncompressible}
\end{equation}
where $\n$ is the outward pointing unit normal on $S$, $\rho$ is the fluid density, $\vel$ is the fluid velocity, $\vel_s$ is the velocity of surface $S$, and $\gvec{\tau}$ is the viscous stress tensor which together with the thermodynamic pressure $p^\prime$ makes the Cauchy stress tensor 
$\vec{T}=-p^\prime\I+\gvec{\tau}$.
The dynamic pressure in Eq. (\ref{eq:momentumCLIncompressible}) is defined as
\begin{equation}
    p = p^\prime - \rho\vec{g}\cdot\vec{r},
\end{equation} 
where $\vec{g}$ is the gravitational acceleration vector and $\vec{r}$ is the position vector. 
Furthermore, the viscous stress tensor of the Newtonian fluid is given by
\begin{equation}
    \gvec{\tau} =  \eta (\nabla\vel + \nabla\vel^{T}),
\end{equation}
where $\eta$ is the dynamic viscosity.

%\todo[inline]{@TM/ZT: absolute pressure\\Dieters remark: what is this? how is it defined? My input for an answer: the momentum balance involves contact forces (in addition to body forces) and they are given as T.n, where T is the Cauchy stress tensor (or, just "stress tensor"). T can be decomposed into T = -p I + tau (the notation from the paper, using tau as the viscous part of T), where tau is the traceless part of T. This latter property defines p, viz. p is -1/3 trace(T). That's a clear definition of p. Then, by the Newtonian closure law, tau = ... as in the manuscript. And then p is modified by subtracting...\\~ \\MS: I would be Ok with a somewhat shorter definition.}

The relationship between the rate of change of the volume $V$ and the velocity
$\vel_s$ is defined by the {\it {geometric (space)} conservation law}
(GCL, see \cite{thomas:GCL,demirdzicPeric:SCL}):
\begin{equation}
  \ddt{}\int\limits_{V}\dV
  - \oint\limits_{S}\vel_s \cdot \n \,\dS = 0.
  \label{spaceCL}
\end{equation}
The formulation of the above mathematical model is a well-known ALE formulation. % ALE already introduced

\subsection{Interface coupling conditions}
%\label{secSub:??}

If fluid phases are immiscible, fluid flow \cref{eq:massCLIncompressible,eq:momentumCLIncompressible} can be applied
for each phase separately, while on the interface the proper boundary conditions
must be used.
The relation between fluid velocities on the two sides of the interface is
determined by the {\it kinematic condition},
which states that the velocity must be continuous across the interface:
\begin{equation}
  \vel_{\rm l} = \vel_{\rm g},
  \label{eq:KinCond}
\end{equation}  
where $\vel_{\rm l}$ and $\vel_{\rm g}$ are the fluid velocities at the two sides (e.g.\ liquid and gas) of the fluid interface. 

The dynamic condition follows from the momentum conservation law and states that forces acting on the fluid at the interface are in equilibrium:
\begin{equation}
    (p_{\rm g} - p_{\rm l})\n_\Sigma - 
    (\gvec{\tau}_{\rm g} - \gvec{\tau}_{\rm l}) \cdot \n_\Sigma = 
    \sigma \kappa \n_\Sigma + \sGrad\sigma - 
    (\rho_{\rm g} - \rho_{\rm l})\,(\vec{g}\cdot\vec{r})\,\n_\Sigma,
    \label{eq:InterForceBalance}
\end{equation}
%\todo[inline]{Dieter:\\in standard math notation, I would put \cdot n_Sigma to the right, as I work with column-vectors. One might use row-vector notation as well. Is this meant?\\~\\MS: I put it to the right}
%\todo[inline]{Dieter:\\this way of writing the jump equation is not orientation invariant! either use our standard jump notation [[...]] or add the information, which orientation the normal needs to have in order to have the correct sign (of kappa)! ZT: it is defined that normal }
%
where $\n_\Sigma$ is the unit normal vector on the interface, which points from the liquid phase to the gas phase, $\kappa = - \sGrad \dprod \n_\Sigma$ is twice the mean curvature of the interface, $\sigma$ is the surface tension coefficient and $\sGrad \sigma$ is the gradient of the surface tension coefficient, where $\sGrad = \nabla( \I-\n_\Sigma \n_\Sigma)$ is the surface gradient operator. The last term on the right-hand side of \cref{eq:InterForceBalance} appears due to transformation of the thermodynamic pressure to the dynamic pressure at the interface. The pressure jump across the interface is obtained by taking the normal component of the force balance (\cref{eq:InterForceBalance}):
\begin{equation}
  p_{\rm g} - p_{\rm l} = 
  \sigma \kappa - 2 \left(\eta_{\rm g} - \eta_{\rm l} \right) \sGrad \cdot \vel -
  (\rho_{\rm g} - \rho_{\rm l})\,(\vec{g}\cdot\vec{r}),
  \label{eq:PressureJumpEq}
\end{equation}
The second term on the right-hand side of \cref{eq:PressureJumpEq} represents the jump of the normal viscous force across the interface, expressed through the surface divergence of the interface velocity. For example, the normal viscous force at the interface can be expressed as (see \cite{chenSaricStone:devNormalStress}):
\begin{equation}
(\n_\Sigma \otimes \n_\Sigma):\gvec{\tau} = 
2\eta\,(\n_\Sigma \n_\Sigma):\grad\vel =
-2\eta\sGrad\cdot\vel,
\end{equation}
where the following identity, being valid for an incompressible fluid flow ($\nabla\cdot\vel=0$), is used:
\begin{equation}
    \nabla\cdot\vel = 
    \sGrad\cdot\vel + (\n_\Sigma \otimes \n_\Sigma):\grad\vel.
\end{equation}
By taking the tangential component of the force balance (\cref{eq:InterForceBalance}) one obtains a relation between the normal derivative of the tangential velocity on the two sides of the interface:
\begin{equation}
%\begin{split}
   \eta_{\rm g}\left(\grad\vel_{\rm t}\cdot \n_\Sigma\right)_{\rm g} 
 - \eta_{\rm l}\left(\grad\vel_{\rm t}\cdot \n_\Sigma\right)_{\rm l} =
 - \sGrad \sigma - \left(\eta_{\rm g} - \eta_{\rm l} \right)\left(\sGrad v_{\rm n}\right),
%\end{split}
  \label{eq:NormalVelocityGradientRelationEq}
\end{equation}
where it is taken into account that the tangential component of the viscous force at the interface can be expressed as follows:
\begin{equation}
    (\gvec{\tau}\cdot\n_\Sigma)\cdot(\I-\n_\Sigma \otimes \n_\Sigma) =
    \eta\left(\grad\vel_{\rm t}\cdot\n_\Sigma+\sGrad v_{\rm n}\right),
\end{equation}
where $\vel_{\rm t} = (\I-\n_\Sigma \otimes \n_\Sigma)\cdot\vel$ is the tangential velocity component and $v_{\rm n} = \n_\Sigma\cdot\vel$ is the normal velocity component at the interface. 
A non-zero gradient $\sGrad \sigma$ of the surface tension coefficient  
%
%\todo{Dieter:\\put behind "gradient"\\~\\MS:\\remove word "coefficient"? Then I think the formula is correct here}
%
can occur for example due to a nonuniform distribution of surfactant at the interface or due to the presence of a temperature gradient.

\subsection{Surfactant transport}
%\label{secSub:??}

If impact of surfactant on the flow field is to be analyzed, it is necessary to extend the mathematical model by an equation which governs the transport of surfactant in the liquid bulk and an equation which governs the transport of surfactant along the interface.
%
%\todo{(2x in this sentence:\\Dieter:\\use "surfactant", which is a species (with many molecules...). Otherwise, you refer to several different (types of) surfactants.\\~\\MS:\\Can we use "surfactant(s)"?}
%
The bulk surfactant transport equation in ALE form reads as follows:
\begin{equation} \label{eq:volSurfactTransp}
    \ddt{}\int_V c \dV 
  + \oint_S  (\vel-\vel_s) c \cdot \n \, \dS 
  - \oint_S (D\grad c)\cdot \n \dS = 0,
\end{equation}
where $c$ is the volume-averaged bulk surfactant concentration and $D$ is the bulk surfactant diffusion concentration.

The governing equation for surfactant transport along a surface patch $S$
%
%\todo[inline]{Dieter:\\arbitrary can be misleading. It should have a regular boundary etc. Possibly omit this. In total: arbitrary surface => surface patch\\~\\Surface:\\probably more clear to say "surface patch", meaning a part of the full surface $\Sigma$\\~\\remove this todo, when read!}
%
attached to the fluid interface ($S \subset \Sigma(t)$) and bounded by a closed 
%
%\todo{Dieter:\\what is meant by "moving"? It will move with Sigma(t) in any case. Does it refer to tangential motion? I guess so. As this is very important, it should be explained in more detail!\\~\\MS:explained enough in the rest of the sentence under the eq?} 
%
curve $\partial S$ which can move along the interface reads:
\begin{equation} \label{eq:boundarySurfTransp}
    \ddt{}\int_S \Gamma \dS 
  + \oint_{\partial S} (\vel - \vec{b})\cdot\vec{m}\,\Gamma \dL
  - \oint_{\partial S} (D_\Gamma \sGrad\Gamma)\cdot\vec{m} \dL = \int_S s_\Gamma \dS,
  %\label{eq:surfactantCLSurface}
\end{equation}
where $\Gamma$ is the surfactant concentration at the interface, $\Gamma_\infty$ is the saturated surfactant concentration 
%
%at the interface,
%Dieter: makes not much sense. It will usually be a material property. It can depend on pressure and temperature, hence indirectly on the position, but we should no go into this. => replace by "under the given thermodynamic conditions".
%
under the given thermodynamic conditions, $\vec{m}$ is the outward pointing unit bi-normal on $\partial S$, 
%
%$\vel_t = (\I -\n_\Sigma\n_\Sigma)\cdot\vel$ is the tangential component of fluid velocity at the interface, 
%
$\vec{b}$ is the velocity at which the curve $\partial S$ moves along the interface,
%
%\todo[inline]{Dieter:\\what does this mean "along the interface"? sounds like a tangential part. But then the notation $b_t$ is not needed above. I think it is the velocity of points on the curve in full space, i.e. normal + tangential part. It is also not a property of the curve as a whole, but of points which make up the curve...}
%
$L$ is the arc length measured along $\partial S$, $D_s$ is the diffusion coefficient of the surfactant along the interface and $s_\Gamma$ is the source/sink of surfactants per unit area due to adsorption and desorption.

For the sake of simplicity,
we focus in this paper on the prominent Langmuir’s kinetics law \cite{birdi_chemPhys_collSysInterf_2002,brenner2013interfacial,pesci_experimental_2017_book} 
for the transfer of surfactant between the bulk and the interface due to adsorption and desorption:

\begin{equation}
    s_\Gamma = k_{\rm a}\left[c_{\rm I}(\Gamma_\infty-\Gamma)-\beta\Gamma\right],
\end{equation}
where $k_{\rm a}$ and $\beta$ are the parameters of the adsorption and desorption kinetics, respectively, and $c_{\rm I}$ is the value of the bulk surfactant concentration at the interface. The implementation allows for a straightforward extension to cope with other sorption models. The normal derivative of bulk surfactant concentration at the interface is given by following expression:
%
%\todo[inline]{Dieter on "is defined.."\\for sure NOT!!
%It is of course defined from the concentration field in the bulk! (15) is nothing but mass conservation during transfer between bulk and interfaces!}
%
\begin{equation}
    (\grad c\cdot \n)_{\rm \Sigma} = -\frac{s_\Gamma}{D}.
    \label{eq:surfactInterBC}
\end{equation}
%
%\Cref{eq:surfactInterBC} defines the surfactant diffusion flux at the interface which can be used only if there is adequate mesh resolution next to the interface in normal direction. If this is not the case, an alternative approach is to use the Subgrid-Scale (SGS) model described in \cref{sec:SGS}.
%
In order to account for the surfactant bulk-interface mass transfer balance (\cref{eq:surfactInterBC}), the normal derivative of the surfactant concentration at the interface needs to be calculated with sufficient accuracy. Due to convection along the interface and typically small surfactant diffusivities, extremely thin boundary layers appear and make this calculation a challenging step. At this point, use of \otherChange{Subgrid-Scale} modeling for improved flux calculation as described in \cref{sec:SGS} is beneficial.

The surface tension is related to the surfactant concentration on the interface and given for the Langmuir sorption model with its adsorption isotherm by the following equation of state 
%by the nonlinear Frumkin-Langmuir equation of state \cite{brenner2013interfacial}:

\begin{equation}
    \sigma = 
    \sigma_0 + \mathcal{R}T\Gamma_\infty\ln{\left(1 - \frac{\Gamma}{\Gamma_\infty}\right)},
    \label{eq:surf-tension-surfact}
\end{equation}
where $\sigma_0$ is the surface tension of a clean interface, $\mathcal{R}$ is the universal gas
constant and $T$ is the temperature \cite{birdi_chemPhys_collSysInterf_2002,brenner2013interfacial,pesci_experimental_2017_book}. %Formula 15.34. in pesci2017
\Cref{eq:surf-tension-surfact} can also be written as
\begin{equation}
    \sigma = 
    \sigma_0\left[ 1 + \beta_s\ln{\left(1 - \frac{\Gamma}{\Gamma_\infty}\right)}\right],
    \label{eq:surf-tension-surfact-2}
\end{equation}
where $\beta_s=\frac{\mathcal{R}T\Gamma_\infty}{\sigma_0}$ is the elasticity number.
% For the test case described in \cref{secSub:results-marangoni} we added a linear equation of state:
% %
% \begin{equation}
%     \sigma = 
%     \sigma_0\left( 1 + \beta_s\frac{\Gamma}{\Gamma_\infty}\right).
%     \label{eq:surf-tension-surfact-lin}
% \end{equation}

\section{Finite Volume and Finite Area discretisation and equation-system solution}
\label{sec:ALE-IT}

\subsection{Discretisation of the computational domain}

The cell-centered Finite Volume Method (FVM, \cite{Tukovi2012}) is used to discretise the bulk fluid models (volumetric transport equations) and the face-centered Finite Area Method (FAM, \cite{Tukovi2012}) is used to discretise the surfactant transport model, a transport equation on the fluid interface (moving surface).
\ReviewerTwo{This discretization is sketched in \cref{fig:InterfaceMesh,fig:freeSurfaceMesh} and can be seen for a case with a bubble in \cref{fig:risBubble_mesh}.}
The finite volume/area discretisation is based on the integral form of the conservation equation over a fixed or moving control volume/area. The discretisation procedure is divided into two parts: discretisation of the computational domain and the equation discretisation. Here we provide a brief description of both discretisation methods together with a description of the interface tracking method, while more details can be found in \cite{Tukovi2012,Tukovi2018,tukovic2008simulation}.

\begin{figure}
\centering
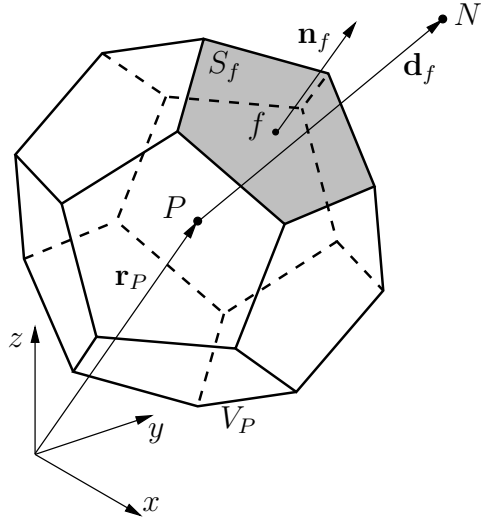
\caption{Polyhedral control volume (cell) and the notation used (adopted from \citep{Tukovi2012})}.
\label{fig:cell}
\end{figure}

The time interval is split into a finite number of time-steps $\Delta t$ and the equations are solved in a time-marching manner. The computational space is divided into a finite number of polyhedral control volumes (CV) or cells bounded by polygons. The cells do not overlap and fill the spatial domain completely. \Cref{fig:cell} shows a polyhedral control volume $V_{P}$ around the computational point~$P$ located in its centroid, face $f$ with area $S_{\!f}$, face unit normal vector $\mathbf{n}_{f}$ and the centroid $N$ of a neighboring CV sharing the face $f$. The geometry of the CV is fully determined by the position of its vertices.

In case the solution domain is changing in time, e.g., when the fluid interface deforms, the ﬁnite volume mesh is adjusted to the time-varying position or shape of the solution domain boundaries using a deforming mesh approach. The internal CV-vertices are moved based on the prescribed motion of the boundary vertices, while the topology of the mesh stays unchanged. In the implemented moving mesh interface tracking solver, a finite-volume automatic mesh motion solver is applied, where the Laplace equation is solved for mesh-point displacements with variable diffusion coefficient, using the polyhedral unstructured cell-centered FVM.

\begin{figure}
\centering
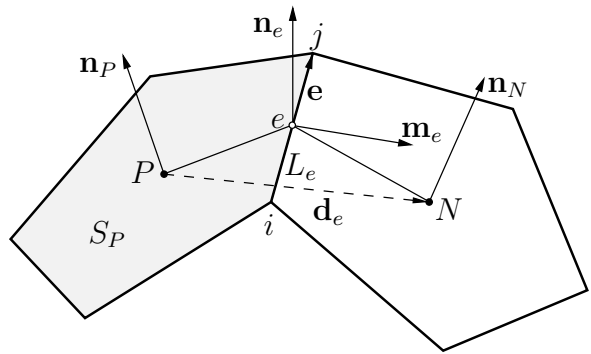
\caption{Polygonal control area.}
\label{fig:finite-area}
\end{figure}

The polygonal cell-faces which coincide with the interface constitute the finite area mesh on which the surfactant transport equation is discretised. \Cref{fig:finite-area} shows a sample polygonal control area $S_{\!P}$ around the computational point $P$ located in its centroid, the edge $e$, the edge length $L_{e}$, the edge unit bi-normal vector $\mathbf{m}_{e}$ and the centroid $N$ of the neighbouring control area sharing the edge $e$. The bi-normal $\mathbf{m}_{e}$ is perpendicular to the edge normal $\n_{e}$ and to the edge vector $\vec{e}$.

\subsection{Discretisation of volumetric equations}

The second-order collocated FV discretisation of an integral conservation equation transforms the surface integrals into sums of face integrals and approximates them and the volume integrals to the second order accuracy. \ReviewerTwo{Error estimates for the unstructured FVM have been analyzed in detail, e.g., in \citep{jasak:PhD,juretic_error_2010}. We rely on second-order divergence and Laplace term discretization and second-order interpolation schemes, denote approximated quantities with $\approx$ and omit the error terms in the description below for clarity, to place focus on the discretization.}
The spatially discretised form of the momentum \cref{eq:momentumCLIncompressible} for the moving control volume $V_{P}$ reads: 

\begin{equation}
  \begin{split}
    &\ddt{\left(\vel_P V_P\right)} +
    \sum_{f} (\dot{V}_{f} - \dot{V}_{s,f})\,\vel_{f}\\
    &=\sum_{f} \nu_{f}\,[(\grad\vel)\cdot\n]_{f}\,S_{f}
    + (\grad p)_{P} V_{P},
  \end{split}
  \label{eq:DiscretMomentumALE}
\end{equation}
where the subscripts ${}_{P} $ and ${}_{f}$ represent the cell-center and face-center values. The volume flow rate due to fluid flow through the cell-face, $\dot{V}^{n+1}_{f} = (\n\cdot\vel S)^{n+1}_{f}$ must satisfy the discretised mass conservation law, where the superscript ${}^{n+1} $ represents the new time-step, while the volume flux due to grid motion, $\dot{V}^{n+1}_{s,f}$, must satisfy the discretised GCL, \cref{spaceCL}. The evaluation of the mesh-face volume fluxes is described in \cite{ferzigerPeric,Tukovi2018}.

The face-center values of all dependent variables are calculated using linear interpolation of the neighbouring cell-center values. The exception is the face value of the dependent variable in the convection term [$\vel_{f}$ in \cref{eq:DiscretMomentumALE}] which must be calculated using some of the bounded convection discretisation schemes available in OpenFOAM.

The face-normal derivative of the velocity $[\n\cdot\grad\vel]_f$ is discretised using the linear scheme with non-orthogonal and skewness correction (see \cref{fig:cell}):
\begin{equation}
    (\grad\vel\cdot \n)_f \ReviewerTwo{\approx}
        \frac{\vel_{N}-\vel_{P}}{d_{{\rm n},f}} +
        \frac{(\grad\vel)_{N}\cdot\vec{k}_{N}-(\grad\vel)_{P}\cdot\vec{k}_{P}}{d_{{\rm n},f}} \,,
  \label{eq:NormalDerivative}
\end{equation}
where $d_{{\rm n},f} = \n_{f}\cdot\vec{d}_{f}$, $\vec{k}_{P} = (\I-\n_{f}\otimes\n_{f})\cdot(\vec{r}_{f} - \vec{r}_{P})$ and $\vec{k}_{N} = (\I-\n_{f}\otimes\n_{f})\cdot(\vec{r}_{N} - \vec{r}_{f})$.
The first term on the right-hand side of \cref{eq:NormalDerivative} is treated implicitly, while the correction term is treated explicitly.

The cell-face volumetric fluid flux in the non-linear convection term is treated explicitly after the discretisation, i.e.\ its values from the previous iteration are used. The cell-center gradient of the dependent variables used for the calculation of the explicit source terms and the non-orthogonal correction in \cref{eq:NormalDerivative}, is obtained by using the least-squares fit \cite{demirdzic:AleFVM}. This method produces a second-order accurate gradient irrespective of the mesh quality.

The {\bf temporal discretisation} of all equations is performed by using the second-order accurate implicit three time level scheme \cite{ferzigerPeric} referred to as the \parameter{backward} scheme in OpenFOAM. All terms of \cref{eq:DiscretMomentumALE} are evaluated at the new time instance $t^{n+1} = t^n+\Delta t$ and the temporal derivative is discretised by using two old-time levels: 
\begin{equation}
    \left(\!\ddt{\vel}\!\right)^{\!{n+1}} \ReviewerTwo{\approx} 
    \frac{3\vel^{n+1} - 4 \vel^n + \vel^{n-1}}{2\Delta t},
    \label{firstDerivativeBackward}
\end{equation}
where $\vel^{n+1} = \vel(t^n+\Delta t)$, $\vel^n = \vel(t^n)$ and $\vel^{n-1} = \vel(t^n-\Delta t)$. One can also use other temporal discretisation schemes available in OpenFOAM.

When \cref{eq:NormalDerivative,firstDerivativeBackward} are substituted into \cref{eq:DiscretMomentumALE} and the convection discretisation scheme is applied, the discretised form of the momentum \cref{eq:momentumCLIncompressible} can be written in the form of a linear algebraic equation, which for cell $P$ reads:
\begin{equation}
  a_{P}\,\vel^{n+1}_{P} + \sum_{N} a_{N}\,\vel^{n+1}_{N} = \vec{r}_{P} + 
  (\grad p)_{P},
  \label{linEqnFluid}
\end{equation}
where the diagonal coefficient $a_{P}$, the off-diagonal coefficients $a_{N}$ and the source term $\vec{r}_{P}$ can be found in \cite{Tukovi2018}.

The mathematical model of fluid flow is solved using a segregated solution procedure where the discretised momentum equation is solved decoupled from the discretised pressure equation. The  discretised pressure Poisson equation, obtained by combining the discretised momentum and continuity equations, reads:
\begin{equation}
    \sum_{f} \left(\frac{1}{a}\right)_{f}
    (\n\cdot\grad p)^{n+1}_f\,S^{n+1}_f
  = \sum_f \n^{n+1}_f\cdot\left(\frac{\vec{H}}{a}\right)_f S^{n+1}_f,
    \label{pressureEqn}
\end{equation}
where the $(1/a)_f$ and $(\vec{H}/a)_f$ terms are calculated by using the temporally consistent Rhie-Chow interpolation procedure proposed in \cite{Tukovi2018}.

The face normal derivative of the pressure at the left hand side of \cref{pressureEqn} is discretised using \cref{eq:NormalDerivative} applied on the pressure field. The absolute volume fluid flux $\dot{V}^{n+1}_{\!f}$ through the cell face~$f$ is calculated as follows:
\begin{equation}
    \dot{V}^{n+1}_f =
    \n^{n+1}_f\cdot\left[\left(\frac{\vec{H}}{a}\right)_f - 
    \left(\frac{1}{a}\right)_f (\grad p)^{n+1}_f\right]S^{n+1}_{\!f}.
    \label{volumeFlux}
\end{equation}
Volume flux calculated in this manner will satisfy the discretised continuity equation if the pressure field satisfies the pressure \cref{pressureEqn}.

\subsection{Discretisation of surface equations}

Applying a second-order collocated finite area method, the surfactant transport \cref{eq:boundarySurfTransp} can be discretised on the moving control area $S_{\!P}$ (see \cref{fig:finite-area}) as follows:
\begin{equation}
%  \begin{split}
    \ddt{\left(\Gamma_{\!P}\,S_{\!P}\right)} +
    \sum_{e} (\vec{m}\cdot \vel)_e L_{e}\,\Gamma_{e}
    =\sum_{e} D_{\Gamma,e} (\vec{m}\cdot\sGrad\Gamma)_{e} L_{e}
    + (s_\Gamma)_{P}\, S_{\!P},
%  \end{split}
  \label{eq:DiscretSurfactant}
\end{equation}
where the subscripts ${}_P$ and ${}_e$ represent the face-center and edge-center values, and it is assumed that finite-area points move in normal direction which means that $(\vec{m}\cdot\vec{b})_{e} = 0$.
%
%\todo{@TM/ZT: b? I don't see it in \cref{eq:DiscretSurfactant} nor in \cref{fig:finite-area}}\todo{Dieter:\\only appears with v. If this is =0, it is strange.}

\begin{figure}
    \centering
    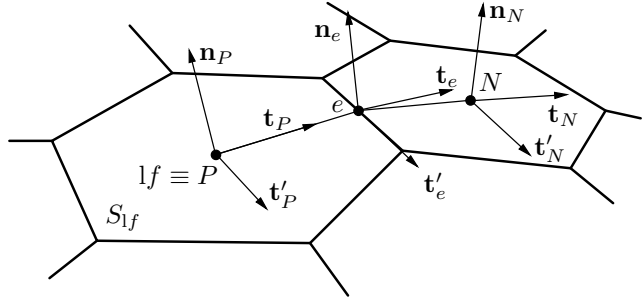
    \caption{Edge-based local orthogonal coordinate system whose axis are aligned with the orthogonal unit vectors $\vec{n}$, $\vec{t}$ and $\vec{t}^\prime$, where vector $\vec{t}$ is tangential to the geodetic line $\overline{PeN}$}
    \label{fig:face-to-edge}
\end{figure}

The edge-center tangential velocity $\vel_{\rm t}$ is calculated using following linear interpolation formula:
\begin{equation}
  \vel_{{\rm t},e} = (\vec{T}_{e})^{\rm T}\cdot
  \left[e_x \vec{T}_{P}\cdot\vel_{{\rm t},P} + 
  (1-e_x)\vec{T}_{N}\cdot\V_{{\rm t},N}\right].
  \label{vectorInterpolFa}
\end{equation}
where $e_x$ is the interpolation factor which is calculated as the ratio of
the geodetic distances $\overline{eN}$ and $\overline{PeN}$ (see \cref{fig:finite-area}):
\begin{equation}
  e_x = \frac{\overline{eN}}{\overline{PeN}},
\end{equation}
and $\vec{T}_{P}$, $\vec{T}_{N}$ and $\vec{T}_{e}$ are the tensors of transformation from the global Cartesian coordinate system to the $(\mathbf{t}^\prime_e,\mathbf{t}_e, \mathbf{n}_e)$ 
%
%\todo{Denote $\mathbf{x}_e$ in Fig 4.\\Tomislav:is now $(\mathbf{t}^\prime_e$} 
%
edge-based local coordinate system defined in \cref{fig:finite-area}. The convection term in \cref{eq:DiscretSurfactant} is discretised using the linear upwind discretisation scheme by taking into account geodetic distances between the neighbouring control area centers. The diffusion term is discretised using the central differencing scheme with non-orthogonal correction \cite{jasak:PhD}:
\begin{equation}
  (\vec{m}\cdot\sGrad\psi)_{e} = 
  \LaTeXunderbrace{\left|\vec{\Delta}_{e}\right| 
  \frac{\psi_{N}-\psi_{P}}{L_{PN}}}_{\textrm{Orthogonal contribution}} +
  \LaTeXunderbrace{\vec{k}_{e}\cdot(\sGrad\psi)_{e},}_{
    \textrm{Non-orthogonal correction}}
  \label{nonOrthoCorrFa}
\end{equation}
% LaTeXunderbrace: https://tex.stackexchange.com/questions/117628/mathabx-mathtools-extremely-odd-underbrace-behaviour-how-to-fix
where $\vec{k}_e = \vec{m}_{e} - \vec{\Delta}_{e}$, $\vec{\Delta}_{e} = \frac{\vec{t}_e}{\vec{t}_{e}\cdot\vec{m}_{e}}$, $L_{PN}$ is the geodetic distance $\overline{LeP}$ and $\vec{t}_{e}$ is the unit tangential vector to the geodetic line $\overline{PeN}$ at the point $e$ (see \cref{fig:face-to-edge}).

\subsection{Interface tracking procedure}

\begin{figure}[hptb]
    \centering
    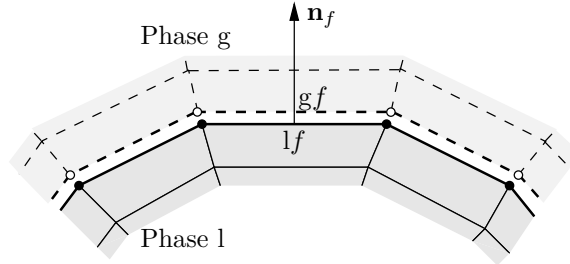  
    \caption{Representation of the interface with the mesh boundary faces.}
    \label{fig:InterfaceMesh}
\end{figure}

The numerical modelling of the two-phase fluid flow with a sharp interface is performed using a moving mesh interface tracking procedure. The computational mesh consists of two separate parts, where each of the meshes covers only one of the considered two fluid phases, see \cref{fig:interfaceTracking}. The two meshes are in contact over two geometrically equal surfaces, $S_{\!\rm l}$ and $S_{\!\rm g}$, at the boundary between the phases, i.e.\ the interface. Surface $S_{\!\rm l}$ represents the liquid side of the interface, and surface $S_{\!\rm g}$ represent the gas side of the interface. Each surface is defined by a set of boundary faces, see \cref{fig:InterfaceMesh}, where each face ${\rm l}f$ on the surface~$S_{\!\rm l}$ has a corresponding geometrically equal face ${\rm g}f$ on the surface~$S_{\!\rm g}$. %TODO: something wrong here
Matching of the two meshes at the interface is assumed in order to make the explanation of the interface tracking method clearer, and is not required in general.  
%For non-matching meshes, a second-order inverse-distance weighted interpolation can be used.

% \begin{figure}[hptb]
%     \centering 
%     \input{./figures/fvm-fam-itm/interfaceTracking.pdf_t} 
%     \caption{Definition of the spatial domain for the moving mesh interface tracking method.}
%     \label{fig:interfaceTracking}
% \end{figure}

Coupling of flow equations between fluid phases is performed by applying adequate boundary conditions at the boundary faces which define the side $\rm l$ and $\rm g$ of the interface. At the side $\rm l$ the pressure $p_{\rm l}$ and normal velocity derivative $(\grad\vel \cdot \n)_{\rm l}$ are specified, while at the side $\rm g$ the velocity $\vel_{\rm g}$ is specified and normal derivative of dynamic pressure is set to zero, $(\n\cdot\grad p)_{\rm g} =0$. The boundary conditions are calculated using the kinematic and dynamic conditions as follows:
\begin{enumerate}
  \item The value of dynamic pressure specified on the face ${\rm l}f$ (see
  \cref{fig:InterfaceMesh}) is calculated from the dynamic pressure at the face ${\rm g}f$ using \cref{eq:PressureJumpEq} as follows:
  \begin{equation}
    \begin{split}
    p_{{\rm l}f} &= p_{{\rm g}f} - (\rho_{\rm l} - \rho_{\rm g})\,\vec{g}\cdot\vec{r}_{{\rm l}f}\\ 
    &- (\sigma \kappa)_{{\rm l}f} - 
    2 \left(\eta_{\rm l} - \eta_{\rm g} \right) (\sGrad \cdot \vel)_{{\rm l}f},
    \end{split}
    \label{pressure_l}
  \end{equation}
  where $p_{{\rm g}f}$ is the dynamic pressure at the face~${\rm g}f$ calculated by extrapolation from the fluid~$\rm g$, and $\vec{r}_{{\rm l}f}$ is the position vector of the face center~${\rm l}f$. The surface divergence of the velocity vector~${(\sGrad \cdot \vel)_{{\rm l}f}}$ at ${\rm l}f$ is calculated using the surface Gauss integral theorem \cite{weatherburn:diffGeo3D}. The procedure for calculating the surface force $(\sigma \kappa)_{{\rm l}f}$ is described later in this section.
  \item The normal velocity derivative specified at ${\rm l}f$ is calculated from the normal velocity gradient at ${\rm g}f$ using \cref{eq:NormalVelocityGradientRelationEq}, as follows:
\begin{equation}
\begin{split}
    \left( \grad \vel \cdot \n \right)_{{\rm l}f} &=
    \frac{\eta_{\rm g}}{\eta_{\rm l}}\,(\I-\n_{{\rm l}f}\otimes\n_{{\rm l}f})\cdot\left[ 
    \left(\grad \vel\right)_{{\rm g}f} \cdot \n_{{\rm l}f} \right]\\
 &+ \frac{1}{\eta_{\rm l}}\,(\sGrad \sigma)_{{\rm l}f} - 
 \n_{{\rm l}f}\,(\sGrad \cdot \vel)_{{\rm l}f}\\
 &+ \frac{(\eta_{\rm g} - \eta_{\rm l})}{\eta_{\rm l}}\left(\sGrad \vel_{\rm n}\right)_{{\rm l}f},
\end{split}
\label{velGrad_1}
\end{equation}
where $\n_{{\rm l}f}=-\n_{{\rm g}f}$ is the unit normal of the face ${\rm l}f$. \Cref{velGrad_1} is derived using the identity $\grad\vel\cdot \n  + \n\,(\sGrad\cdot\vel)=(\I-\n\otimes\n)\cdot(\grad\vel \cdot \n)$. The surface gradient of the normal velocity component $(\sGrad \vel_{\rm n})_{{\rm l}f}$ at ${\rm l}f$ is calculated using the surface Gauss integral theorem. Calculating the tangential surface force $(\sGrad \sigma)_{{\rm l}f}$ is described later in this section.
  \item According to the kinematic condition (\ref{eq:KinCond}), the tangential velocity component specified on ${\rm g}f$ is transferred from ${\rm l}f$:
  \begin{equation}
    (\vel_{\rm t})_{{\rm g}f} = (\vel_{\rm t})_{{\rm l}f}. 
  \end{equation}
  The normal velocity component specified on ${\rm g}f$ is calculated from the condition of zero net mass flux, $(\dot{V}_{{\rm g}f} - \dot{V}_{{\rm g}f})=0$, i.e.\
  \begin{equation}
    (\vel_{\rm n})_{{\rm g}f} = 
    -\frac{\dot{V}_{{\rm g}f}}{S_{{\rm g}f}}\n_{{\rm l}f},
  \end{equation}
where $\dot{V}_{{\rm g}f}$ is the volume flux of the face ${\rm g}f$. Since the displacement of mesh points on the side~$\rm g$ is equal to the displacement on the side $\rm l$ of the interface, $\vec{u}_{{\rm g}i} = \vec{u}_{{\l}i}$, the same is valid for the volume fluxes of the faces ${\rm l}f$ and ${\rm g}f$:
  \begin{equation}
    \dot{V}_{{\rm g}f} = \dot{V}_{{\rm l}f}.
  \end{equation}
\end{enumerate}

\begin{figure}[hptb]
  \centering
  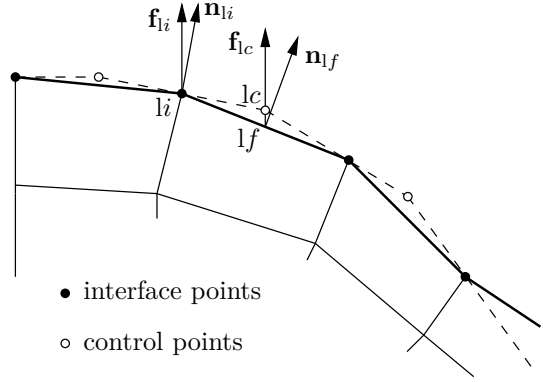  
  \caption{Definition of the interface using control points.}
  \label{fig:ControlPoints}
\end{figure}

The specification of interface boundary conditions using the above procedure is done at the beginning of each outer iteration. In general, at the end of an outer iteration, the net volume flux through the boundary side~$\rm l$ is different from zero, i.e.
\begin{equation}
  [\dot{V}^{\rm p}_{{\rm l}f} - (\dot{V}_s)^{\rm p}_{{\rm l}f}] \neq 0,
\end{equation}
where $\dot{V}^{\rm p}_{{\rm l}f}$ is the fluid volume flux through the face ${\rm l}f$ and $(\dot{V}_s)^{\rm p}_{{\rm l}f}$ is the cell-face volume flux through the same face, both obtained in the previous outer iteration. In order to correct the net volume flux, the interface points must be moved to accomplish the following volume flux corrections:
\begin{equation}
  \dot{V}^\prime_{{\rm l}f} = 
  [\dot{V}^{\rm p}_{{\rm l}f} - (\dot{V}_s)^{\rm p}_{{\rm l}f}].
  \label{meshFluxCorrection}
\end{equation}
The displacement of the interface points is calculated based on the procedure
proposed in \cite{muzaferijaPeric:freeSurfaceFV}, where a control point ${\rm l}c$ is attached to the each face ${\rm l}f$ at the side $\rm l$ of the interface as is shown in \cref{fig:ControlPoints}. The corrected position of the interface points is calculated using the following procedure: 
\begin{enumerate}
  \item Calculate the volume $\delta V^\prime_{{\rm l}f}$ which face ${\rm l}f$ sweeps on the way from the current to the corrected position in order to cancel the net mass flux:
  \begin{equation}
    \delta V^\prime_{{\rm l}f} = 
    \frac{2}{3}\,\dot{V}^\prime_{{\rm l}f}\,\Delta t,
  \end{equation}
  where $\dot{V}^\prime_{{\rm l}f}$ is the volume flux correction for face ${\rm l}f$, \cref{meshFluxCorrection}.
  \item Using the above, the displacement of control points in the direction $\vec{f}_{{\rm l}c}$ is:
  \begin{equation}
    h^\prime_{{\rm l}c} = \frac{\delta V^\prime_{{\rm l}f}}
    {S^{\rm p}_{{\rm l}f}\,\n^{\rm p}_{{\rm l}f}\cdot\vec{f}_{{\rm l}c}},
    \label{eq:CtrlPointDisplacement}
  \end{equation}
  where $S^{\rm p}_{{\rm l}f}$ and $\n^{\rm p}_{{\rm l}f}$ are the area and unit normal of the face ${\rm l}f$ in the previous iteration and $\vec{f}_{{\rm l}c}$ is the control point displacement direction. The new corrected positions of the control points are calculated according to the following expression:
  \begin{equation}
    \vec{r}^n_{{\rm l}c} = \vec{r}^{\rm p}_{{\rm l}c} + 
    h^\prime_{{\rm l}c}\,\vec{f}_{{\rm l}c},
  \end{equation}  
  where $\vec{r}^{\rm p}_{{\rm l}c}$ is the position vector of the control point ${\rm l}c$ before the correction. 
  \item The new position of the interface mesh point ${\rm l}i$ is obtained by projection to the plane which is laid over the corresponding control points using the least square method. 
\end{enumerate}

\subsection{Calculation of interface curvature and surface tension}

Regardless of the approach used for tracking the interface between the
phases in a multiphase fluid flow, the implementation of surface tension is
always demanding.
%
%Unphysical fluid flow called "parasitic currents" will arise around the interface as a consequence of inaccuracy in the calculation of surface tension forces. Parasitic currents may be so strong to destroy the interface and break the calculation.

%The total surface tension force on a closed surface must be zero. This condition has served as a starting point for deriving a procedure for the calculation of surface tension forces.
%\todo[inline]{Dieter:\\??? I don't understand this!surface tension appears in the interfacial momentum balance. This originates in an integral form and this form gives the line integral which is to be computed. No need for this statement! OK?\\~\\MS: maybe add: ...leads to the procedure described in the following?}

\begin{figure}[hptb]
  \centering
  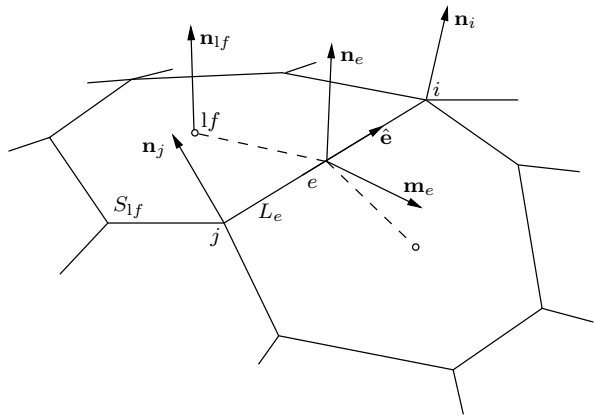  
  \caption{Control area $S_{{\rm l}f}$ at the interface.}
  \label{fig:freeSurfaceMesh}
\end{figure}

Let us assume the interface is discretised with an unstructured surface mesh consisting of arbitrary polygonal control areas. The surface tension force acting on the control area $S_{\!\rm lf}$ (see
\cref{fig:freeSurfaceMesh}) can be expressed by the following equation:
\begin{equation}
  \vec{F}_{{\rm l}f}^\sigma = 
  \oint\limits_{\partial S_{{\rm l}f}}  \sigma \vec{m} \, \dL = 
  \sum_{e} \int\limits_{L_e} \sigma\vec{m}\,\dL
  =\sum_{e} \sigma_{e} \vec{m}_{e} L_{e},
  \label{finiteAreaForce}
\end{equation}
where $\sigma_{e}$ and $\vec{m}_{e}$ are the surface tension and bi-normal unit vector at the center of the edge $e$ and $L_{e}$ is the length of the edge $e$. If the total surface tension force for each control area in the mesh is calculated using \cref{finiteAreaForce}, then the total surface tension force for a closed surface will be exactly zero if the unit bi-normals $\vec{m}_{e}$ for two control areas sharing the edge $e$ are parallel and have opposite direction.

It remains to decompose the surface tension force $\vec{F}_{{\rm l}f}^\sigma$ into the tangential component $(\sGrad\sigma)_{{\rm l}f}$ used in \cref{velGrad_1}, and the normal component $(\kappa\sigma)_{{\rm l}f}\n_{{\rm l}f}$ used in \cref{pressure_l}. Using the surface Gauss integral theorem, the surface tension force acting on the control area $S_{\!{\rm l}f}$ can be expressed by the following equation:
\begin{equation}
  \vec{F}_{{\rm l}f}^\sigma = 
  \int_{S_{{\rm l}f}} \sGrad \sigma \, \dS + \int_{S_{{\rm l}f}} \kappa\sigma\n \, \dS.
  \label{finiteAreaForce_2}
\end{equation}
When the right hand side of \cref{finiteAreaForce_2} is approximated by the mid-point role and the result of the discretisation is equalised with the right hand side of \cref{finiteAreaForce}, the following expression is obtained:
\begin{equation}
  (\sGrad \sigma)_{{\rm l}f} + (\kappa\sigma)_{{\rm l}f}\,\n_{{\rm l}f} = 
  \frac{1}{S_{{\rm l}f}}\sum_{e} \sigma_{e} \vec{m}_{e} L_{e}.
  \label{finiteAreaForce_3}
\end{equation}
Hence, the tangential component of the surface tension force acting on the control area $S_{\!{\rm l}f}$ is equal to the tangential component of the right hand side of \cref{finiteAreaForce_3}:
\begin{equation}
  (\sGrad \sigma)_{{\rm l}f}= \frac{1}{S_{{\rm l}f}}\,(\I-\n_{{\rm l}f}\n_{{\rm l}f})\cdot\sum_{e} \sigma_{e} \vec{m}_{e} L_{e}
  \label{finiteAreaForce_tangent}
\end{equation}
and its normal component is equal to the respective normal component:
\begin{equation}
  (\kappa\sigma)_{{\rm l}f}\n_{{\rm l}f} = 
  \frac{1}{S_{{\rm l}f}}\,(\n_{{\rm l}f}\otimes\n_{{\rm l}f})\cdot\sum_{e} \sigma_{e} \vec{m}_{e} L_{e}.
  \label{finiteAreaForce_normal}
\end{equation}
If the surface tension coefficient is constant ($\sigma=\textrm{const.}$), \cref{finiteAreaForce_tangent} will give a tangential component of the surface tension force equal to zero if the normal unit vector of the control area $S_{\!\rm {\rm l}f}$ satisfies the following equation:
\begin{equation}
  \kappa_{{\rm l}f}\n_{{\rm l}f} = 
  \frac{1}{S_{{\rm l}f}}\sum_{e} \vec{m}_{e} L_{e}
  \label{finiteAreaCurvatureVec}
\end{equation}
or, if $\kappa_{{\rm l}f}\neq 0$:
\begin{equation} 
  \n_{{\rm l}f} = \frac{\sum\limits_{e} \vec{m}_{e} L_{e}}{|\sum\limits_{e} \vec{m}_{e} L_{e}|}.
  \label{finiteAreaNormal}
\end{equation}

With \cref{finiteAreaForce_tangent,finiteAreaForce_normal,finiteAreaCurvatureVec} we shall formulate a procedure for the calculation of the surface tension force which ensures that the total surface tension force on a closed surface will be exactly zero. Unfortunately, the fulfilment of this condition is not sufficient for successful application of surface tension forces in the calculation. Specifically, unphysical fluid flow near the interface arises due to local (rather than global) inaccuracy in the calculation of surface tension forces.

From \cref{finiteAreaForce} one can see that the accuracy of surface tension force calculation depends on the accuracy of calculation of the bi-normal unit vector $\vec{m}_{e}$ which is calculated using the following expression:
\begin{equation}
  \vec{m}_{e} =  \hat{\vec{e}}\times\frac{\n_{i}+\n_{j}}{2},
  \label{m_e_2}
\end{equation}
where $\hat{\vec{e}}$ is the unit vector parallel with edge $e$ and $\n_{i}$ and
$\n_j$ are the interface normal unit vectors in points $i$ and $j$ (see \cref{fig:freeSurfaceMesh}). Using \cref{finiteAreaCurvatureVec,m_e_2} one obtains the exact value of curvature of the control area $S_{\!{\rm l}f}$ if the points of the control area lie on the surface of the sphere.

\subsection{Solution procedure}

Based on the described interface tracking method, one can now define the solution procedure for the Navier-Stokes system on a moving mesh, which may be used for simulating two-phase fluid flow using a moving mesh interface tracking method. The procedure consists of the following steps:
\begin{enumerate}
  \item For the new time instance $t = t^{n+1}$, initialize the values of all dependent variables with the corresponding values from the previous time instance;
  \item Define the displacement directions for the interface mesh points and the control points;
  \item Start of outer iteration loop:
  \begin{enumerate}
    \item Update pressure and velocity boundary conditions at the interface;
    \item Assemble and solve the discretised momentum equation \cref{eq:DiscretMomentumALE} on the mesh with the current shape of the interface. The pressure field, face mass fluxes and volume
    fluxes are used from the previous (outer) iteration;
    \item The velocity field obtained in the previous step is used for the assembly of the discretised pressure \cref{pressureEqn}.
    After the pressure equation is solved, new absolute mass fluxes through the cell faces are calculated. The net mass flux through an interface is generally different from zero;
    \item In order to compensate the net mass flux obtained in the previous step, the interface displacement is calculated using the interface points displacement procedure defined in the previous section;
    \item The interface points displacement is used as a boundary condition for the solution of the mesh motion problem. After mesh displacement, the new face volume fluxes are calculated using the current points positions and the position from the previous time instance; 
    \item Convergence is checked and if the residual levels and the net mass flux through the interface do not satisfy the prescribed accuracy, the procedure is returned to step (a).
  \end{enumerate}
  \item If the final time instance is not reached, return to step 1.
\end{enumerate}

\section{Subgrid-Scale Model}
\label{sec:SGS}

\newcommand{\fcf}{first cell faces normal to the boundary}

\subsection{Motivation}

The simulation of realistic gas-liquid systems is still a huge challenge, e.g.\ when one would like to simulate an industrial scale bubble column reactor. The nature of this application is multi-scale.
One of the smallest scales occurring is that of concentration boundary layers of the rising bubbles at realistic, i.e.\ high, Schmidt numbers.
With conventional FV-methods these scales have to be resolved, what is still unfeasible for an application as the one mentioned above.
This motivated the first paper to discuss SGS modeling \cite{Alke_DNS-VoF_2010}.
In SGS modeling, an analytical function is used to describe the profile of a scalar field inside its boundary layer % it is not the "usual" (velocity) boundary layer!
%a passive scalar boundary layer, a concentration boundary layer in this case,
and to use this SGS information for improved flux computation.

The method has been developed further since and has been applied in the context of Volume of Fluid \cite{weiner_advanced_2017,grunding_reactive_SGS_2016,bothe_fleckenstein_VoF-SGS_2013} as well as in interface tracking frameworks \cite{pesci_experimental_2017_book,pesci_SGS_risBubb_surfactants_2018,Weber2017}.
An overview on the state of \otherChange{Subgrid-Scale} modeling was also given by \cite{weiner_computing_2021}.
Recently, the approach was taken up in \cite{grosso_thermal_2024}, where local curvature and tangential convection effects have been included.

\subsection{Analytical model} \label{secSub:SGS_anaModel}

\begin{figure}[!htbp]
    \centering
    \includegraphics[width=0.5\textwidth]{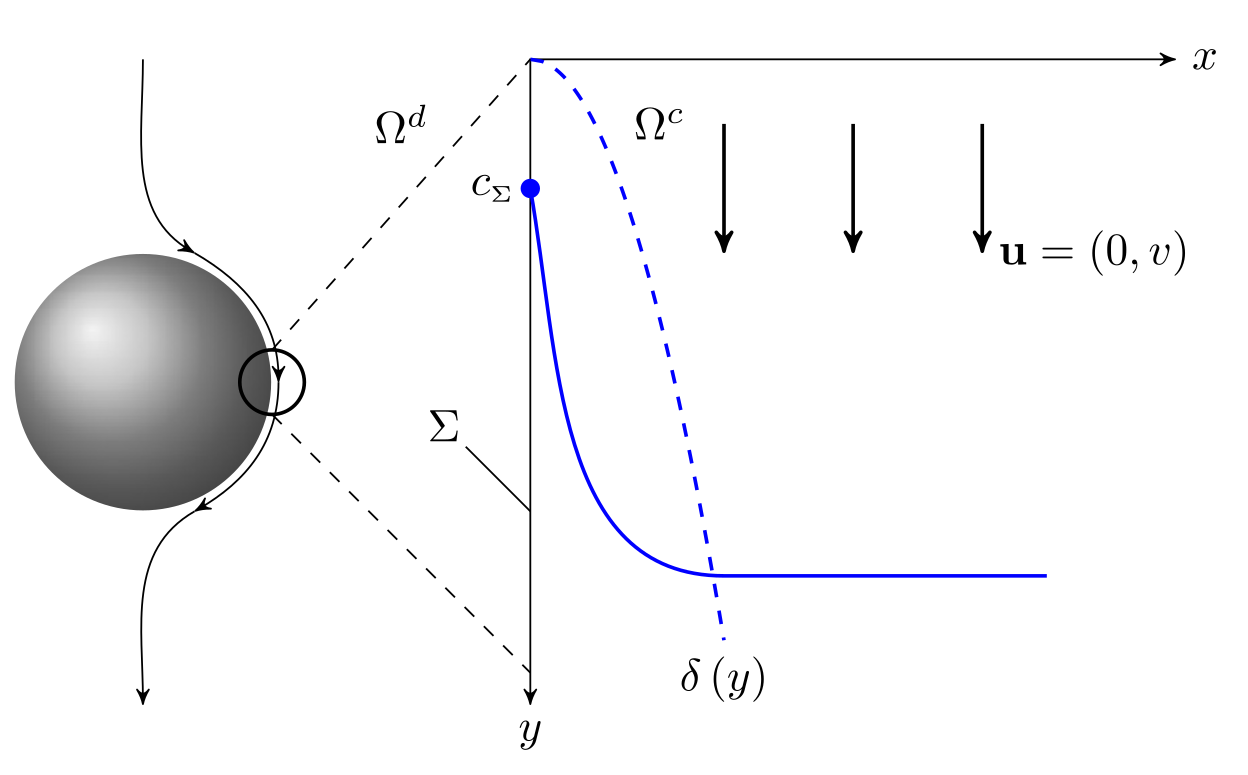}
    \caption{The simplified model for species transfer across a bubble surface. \cite{weiner_advanced_2017} adapted from \cite{bothe_fleckenstein_VoF-SGS_2013}}
    \label{fig:sgs_simplModel}
\end{figure}

% \todo{perfect slip wouldn't hold for the fluid outside a droplet}
The SGS model we implement is based upon analytical solution of a simplified problem, depicted for a bubble in \cref{fig:sgs_simplModel}.
The idea behind the SGS modeling is to zoom in on the interface, so that the curvature gets negligibly small, allowing the analytical model therefore to assume a (locally) flat boundary.
This assumption is adequate for scenarios with a concentration boundary layer thickness small in comparison to the radius of curvature of the given boundary.

At the interface $ \Sigma $ of the model problem (see \cref{fig:sgs_simplModel}), a constant and homogeneous concentration~$ c_{\Sigma} $ is prescribed, which can be computed from Henry's law under the sensible assumption of a well-mixed gas-side concentration, as is pointed out by \cite{bothe_fleckenstein_VoF-SGS_2013}.

The velocity in the substitute model problem is parallel to the interface.
No velocity gradient normal to the boundary exists in the model problem ($ {\partial v}/{\partial x} = 0 $).
There could be a non-zero gradient in the boundary-parallel direction, in which case the boundary layer thickness~$ \delta $ in \cref{eq:SGS_anSol} would change accordingly. %\todo{MS:\\considering y/v as time}
There is no velocity component normal to the boundary.

In the application within ALE-IT, the boundary geometry and the velocity field have to be resolved. % and the interface neighbouring cell’s velocity should be nearly identical to the interface and the next cells velocity.
In a co-moving reference frame, the interface neighbouring cell’s velocity should be nearly parallel to the interface and with the assumption of zero interface-normal velocity gradient the next cells velocity should be nearly identical to it.
When a thin scalar boundary layer exists, the SGS can model the scalar fluxes normal to the boundary without resolving the scalar boundary layer itself.

Further assumptions of the substitute problem are a given, constant value (e.g.\ zero concentration) upstream ($ y < 0 $) of the domain and far away from the boundary ($ x \to \infty $) as well as negligible diffusion in streamwise direction.
%\todo[inline]{Dieter reg. "include":\\so, there are more - which are not told? NOT GOOD!\\~\\MS: yes, like no density change/no gravity, no viscosity change, constant diffusivity coefficient, no limitation by speed of sound/diffusion/light. There is the singularity at y=0. Some unconscious "obvious" assumptions, we are not even aware of...\\Certainly a possible review criticism.}
We also assume quasi-steadiness and the transported scalar being passively transported.

\begin{sloppypar}
Applying these assumptions to the advection-diffusion \cref{eq:volSurfactTransp}
%
% \begin{equation} \label{eq:transport_2} % Weiner2017
%     \partial_t c + \nabla \cdot (\vel c - D\nabla c) = 0
% \end{equation}
% %
% (TODO: equations on interface don't relate to SGS, for derivation of SGS better use eq. from Weiner \ref{eq:transport_2}?)
% \begin{subequations} \label{eq:transport} % Pesci2018, 2.7 & 2.8
%     \begin{alignat}{7}
%         \partial_t c 
%         &+ \nabla 
%         &&\cdot (c \vel &+ \boldsymbol{j}) 
%         &= 0 \qquad &&\text{in } \Omega\setminus\Sigma(t) , \\ %\label
%         \partial_t^{\Sigma} c^{\Sigma} 
%         &+ \nabla_{\Sigma} 
%         &&\cdot (c^{\Sigma} \vel^{\Sigma} &+ \boldsymbol{j}^{\Sigma}) 
%         &= s^{\Sigma} \qquad &&\text{on } \Sigma(t) %\label{}
%     \end{alignat}
% \end{subequations}
with the sorption term ${ s^\Sigma + [\![ \boldsymbol{j} \cdot  \boldsymbol{n}_\Sigma ]\!] = 0 \text{ on } \Sigma(t) }$ %Pesci2018 2.9
and Fick's law $ \boldsymbol{j} = - D \nabla c $ and $ \nu $ being the boundary-parallel velocity, lead to the expression
\end{sloppypar}
%\begin{subequations}
    \begin{equation} \label{eq:sgs_simplEq}
        \nu \frac{\partial c}{\partial y} = D \frac{\partial^2 c}{\partial x^2} \quad \text{for $ x > 0 $ and $ y > 0 $}
    \end{equation}
    with the boundary conditions
    \begin{equation*}
        c(x,y=0) = c_\infty \text{,} \quad c(x\to \infty, y > 0)=c_\infty \quad \text{and} \quad c(x=0,y>0)=c_{|\Sigma}.
    \end{equation*}
%\end{subequations}
The analytical solution to this problem is
\begin{equation}
    \label{eq:SGS_anSol}
    c(x,y)=c_{|\Sigma} + (c_\infty - c_{|\Sigma}) \cdot {\rm erf} \left( \frac{x}{\delta(y)} \right) \qquad \text{with } \delta(y)= \sqrt{4Dy/\nu} .
\end{equation}
The boundary layer thickness~$\delta$ is considered a free parameter, which in IT will be fitted for each interface-attached cell individually as described in \cref{secSubSub:CompDelta}.
The fitting selects the locally best approximation from the family of given error-functions in \cref{eq:SGS_anSol}.

\begin{figure}[!htbp]
    \centering
    \includegraphics[width=0.5\textwidth]{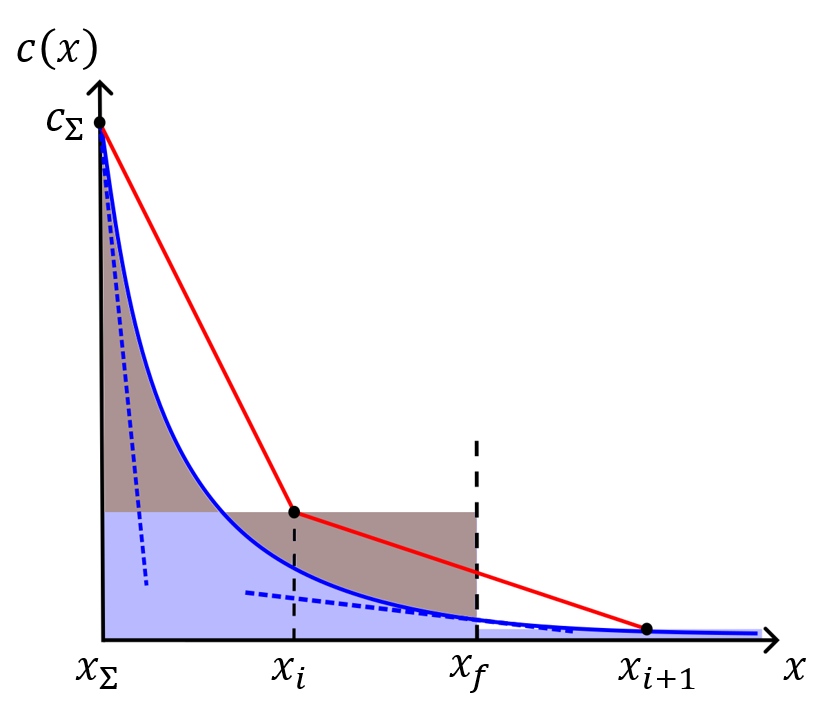}
    \caption{The concentration profile and linear/corrected gradient next to a boundary. Adapted from \cite{weiner_advanced_2017}.}
    \label{fig:sgs_concProfile}
\end{figure}

From the analytical solution for the fitted parameter, we can approximate a more accurate boundary-normal gradient of the scalar concentration at the interface as well as at the \fcf, $ x_f $ in \cref{fig:sgs_concProfile}.
These gradients are used to scale the convective and diffusive scalar fluxes in the Finite Volume discretization, keeping the discretization of convective and diffusive operators implicit.
Scaling is only applied in interface normal direction and described in detail in \cref{secSubSub:CompDiffFluxes,secSubSub:CompAdvFluxes}.

Without the SGS, the maximally second-order unstructured Finite Volume convective and diffusive fluxes are underestimated at the interface and overestimated at the \fcf. \Cref{fig:sgs_concProfile} visualizes the linear gradient calculation in red and the actual gradients at the interface and the \fcf { }with the dashed blue lines.

Note that in the algorithm the concentration in the first cell is not altered by the SGS.
Instead the gradients are used to scale the diffusivities at the respective locations, as described in \cref{secSubSub:CompDiffFluxes,secSubSub:CompAdvFluxes}.

Details of the SGS model can be found in \citep{bothe_fleckenstein_VoF-SGS_2013,weiner_advanced_2017,pesci_SGS_risBubb_surfactants_2018}.
It has been applied to VoF by \cite{weiner_advanced_2017}, where the implementation becomes more complicated.

\subsection{Algorithm} \label{secSub:SGS_algo}

\subsubsection{Equation discretization in finite volume solvers}

The species transport \cref{eq:volSurfactTransp} is discretised using the unstructured Finite Volume Method, e.g.\, using the backward scheme as described above for Navier-Stokes equations,
\begin{equation} %\label{eq:discrTranspEq} %Pesci2018 3.2
    \frac{3 c_P^{n+1} V_P^{n+1} - 4 c_P^n V_P^n + c_P^{n-1} V_P^{n-1}}{\Delta t} + \sum_f \Phi_f c_f^{n+1} = \sum_f D_f (\nabla c)_f^{n+1} \cdot \boldsymbol{S}_f ,
\end{equation}
where $ V_P $ is a control volume/cell, $ \Phi_f = \boldsymbol{S}_f \cdot (\vel - \boldsymbol{w})_f $ is the face flux and the superscripts $ {}^{n+1}, {}^n \text{ and } {}^{n-1} $ symbolize the new, current and preceding timestep. The formula is written for a constant time step size. The subscript $ {}_P \text{ and } {}_f $ represent the value at the cell centers, respectively the face centers. 

For the SGS model to take effect in the computation, we will modify the diffusivity coefficient and the face fluxes in the equation above.

\subsubsection{Computation of model parameter boundary layer thickness} \label{secSubSub:CompDelta}

The model parameter boundary layer thickness $ \delta $ is computed by iteratively comparing the amount of the scalar in the first cell adjacent to the boundary in the simulation, $ \overline{\eta}_c $, with the amount that would be contained in said cell for a given $ \delta $, $\eta_{SGS}$. In \cref{fig:sgs_concProfile} both light brown areas have to be the same size or the area enclosed by the real concentration profile and the cell centered FV-value has to be the same. 
Mathematically this is being expressed in this expression:
\begin{align} \label{eq:cellContentsEqual}
    \overline{\eta}_c = \frac{\overline{c} - c_{|\Sigma}}{c_\infty - c_{|\Sigma}}
    \overset{!}{=}
    \frac{1}{V} \int_V \eta(x/\delta) \,dV = \eta_{SGS}
    \intertext{with } \eta(x,y) = \frac{c(x,y) - c_{|\Sigma}}{c_\infty - c_{|\Sigma}} = {\rm erf} (x/\delta(y)).
\end{align} 

Using the derivative of $ \eta_{SGS} $ with respect to $ \delta $, as described in \cite{pesci_SGS_risBubb_surfactants_2018}, we use a Newton-bisection method until \cref{eq:cellContentsEqual} is fulfilled within a given tolerance.
For details like the initial value and the algorithm in detail the reader is referred to the previously cited article.
\otherChange{We ensure numerical robustness of the Newton-Bisection root finding approach by limiting the next value during the search to a value larger than $2^{-52}$, a floating point number \parameter{SMALL} (machine epsilon) used in OpenFOAM to stabilize floating-point arithmetic. While we recommend this to be always used, we have implemented it as an optional switch.}
For both, the advective and diffusive flux, the same idea is utilized: the scaling of the fluxes according to the boundary layer thickness and therefore the theoretical values given by the SGS.

\subsubsection{Computation of diffusive fluxes}
\label{secSubSub:CompDiffFluxes}

\otherChange{With $ D_f $ being the face-centered molecular diffusivity, $ S_f $ being the area magnitude of the finite volume face and $ \partial_n c $ being the surface normal gradient of the scalar c, the face-centered diffusive flux $ F_f^D $ is given by
\begin{equation}
    F_f^D = -D_f S_f (\partial_n c)_f .
\end{equation}}
In the following, we use the superscript ${ }^{SGS} $ to indicate the value is obtained from, or modified by, the SGS model. The SGS model solves the analytical model for the actual surface normal gradient of $c$, which will be $ (\partial_n c)_f^{SGS} $. However,  \otherChange{we solve the transport equation for $c$ implicitly to retain a high-degree of numerical stability and scale the  diffusion coefficient to account for the influence of the SGS model}. Therefore, we state that the flux computed with the \otherChange{molecular (physical) diffusivity coefficient and the normal gradient from the SGS model is equal to the flux computed from the implicit unstructured FVM gradient and a scaled diffusion coefficient, i.e.} meaning
\begin{subequations}
    \begin{equation}
        F_f^D = -D_f S_f (\partial_n c)_f^{SGS} \overset{!}{=} D^{SGS} S_f  (\partial_n c)_f^o,   
    \end{equation}
    which can be easily re-arranged into
    \begin{equation}
        D_f^{SGS} =  \frac{D_f (\partial_n c)_f}{(\partial_n c)^{SGS}}. 
    \end{equation}
\end{subequations}

This scaling is applied to the cell faces at the boundary as well as at \fcf. For both respective locations the diffusivity parameter needs to be modified at the corresponding position, $ x_{\Sigma} $ and $ x_f $ in \cref{fig:sgs_concProfile}.
There are exceptions, when the correction is not employed, the details of this can be found in \cite{pesci_SGS_risBubb_surfactants_2018}.

\subsubsection{Computation of advective fluxes}
\label{secSubSub:CompAdvFluxes}

In similar manner to the diffusion, the advective flux is given by
\begin{equation} \label{eq:advFlux}
    F^c_{\!f} := c_{\!f}^{SGS} F_{\!f} = c_{\!f} F_{\!f}^{SGS}  \Phi , %TODO: \cdot S ?
\end{equation}
where $ F^c_{\!f} $ is the volumetric flux of the concentration $c$ at the face $f$, and $F_{\!f}:=\mathbf{v}_{\!f} \cdot \mathbf{S}_f$ is the volumetric flux.
Since no convection of the species normal to the interface can take place, the scaling for advective fluxes is only applied to the \fcf. \Cref{eq:advFlux} analogously to our procedure for the diffusive flux, leads to
\begin{equation}
    \Phi^{SGS} = \frac{c^{SGS}}{c} \Phi .
\end{equation}
Exception handling is done similar to diffusion fluxes, of which the details can be found in \cite{pesci_SGS_risBubb_surfactants_2018}.

\section{Software design of the SGS model library} 
\label{sec:SWDesign}

%\subsection{Software design }

The software design is closely related to the model as described in \cref{sec:SGS}.  

% No need to say again what the model does.
% The (TODO: inner workings/sub-?) algorithm, as described in \ref{secSub:SGS_algo}, works in the same way, as in \cite{pesci_SGS_risBubb_surfactants_2018}: we compute the boundary layer thickness and then scale the diffusivity and the concentration, to mimic the behaviour of the given system according to the model of the SGS. 

\begin{figure}[!htbp]
    \centering
    \includegraphics[width=0.7\textwidth]{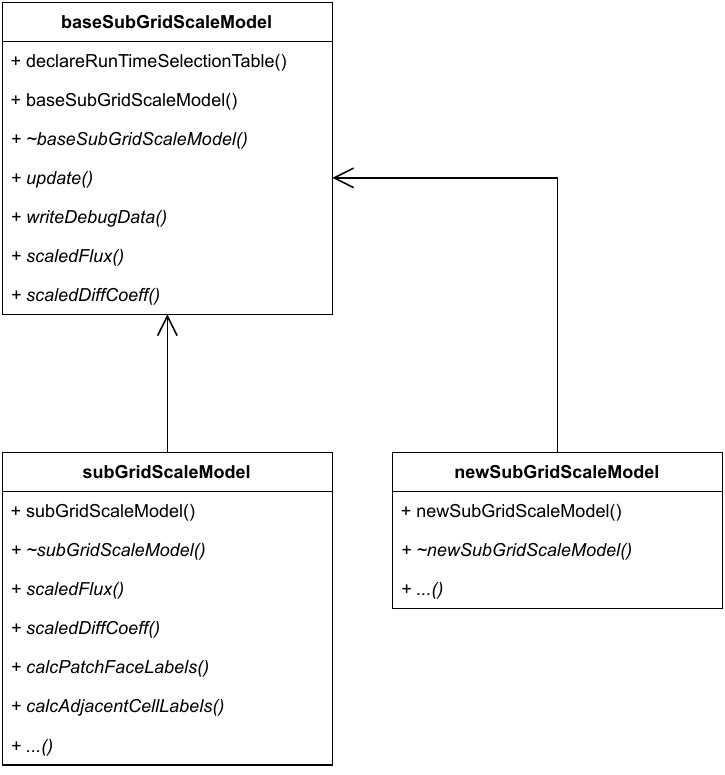}
    \caption{Class diagram of the Subgrid-Scale model, also showing a potential new model}
    \label{fig:SGS_class_UML}
\end{figure}

The first benefit of our implementation is a strict adherence to the OpenFOAM design principles, which enable easy library extension with new models alternative to the SGS \citep{pesci_SGS_risBubb_surfactants_2018} and their Run-Time Selection (RTS) \citep{Maric2014} using configuration files. The implementation is depicted in \cref{fig:SGS_class_UML}, already showing how a new model \parameter{newSubGridScaleModel} would be implemented.

The second benefit of our software design is the possibility to apply SGS models to any Partial Differential Equation (PDE) discretized in OpenFOAM, without requiring extensive modification of existing solvers. Our SGS model library can be applied to any part of the domain boundary with a minimal requirement of a prismatic boundary-adjacent mesh with at least three layers of cells in the boundary layer.

% In contrast to previous approaches the strategy employed uses a temporary field of the modified values in the solution of the scalar transport equation. This is then used by the solver. (TODO: Chapter/Section?)  shows how the model can be applied to an OpenFOAM solver and case.

Our design of the SGS model implements an OpenFOAM library with a class hierarchy and RTS \citep{Maric2014}. A base class which does nothing is present, so an option remains of not using SGS modeling. The SGS model is implemented in a derived class, with virtual functions implementing the functionality from \citep{pesci_SGS_risBubb_surfactants_2018}. All functionality needed by the SGS is either accessed from core OpenFOAM libraries or is contained within the SGS model implementation, which ensures the modularity of the SGS model. An alternative SGS model will require similar addressing and geometric mesh information in the boundary layer and the implementation provided in the original SGS model can be re-used.

To enable the user of the SGS a good understanding of the effects of the SGS model and to make accessible insights for further developments of the SGS model, we implemented the possibility to output various SGS model parameters into OpenFOAM \parameter{volScalarFields}. These include the model parameter~$\delta$ or an indication of the exception handling, what represents the reason the SGS model might not be employed.
These can easily be inspected visually, an example is given in \cref{fig:sgsWedge_Dcorr}, where the diffusion coefficient after the SGS scaling is visualized. One can see, that on the inflow side, the scaling is strong and of opposite effect for the boundary faces versus the faces between adjacent cell and the next outer cell layer. On the outflow side in contrast, there is little effect to be seen. The third layer represents the un-modified diffusion coefficient. Visualizing multiple operating figures of the SGS enables insights into and deepens the understanding of the simulation case and the effect of the SGS to it. \label{text:visualizeFigs}

% What has to be done, to deploy the SGS model is to call the SGS model library in the solver, when the passive scalar equations are being solved and providing the needed model input parameters in the dictionary files of the case being computed.

An exemplified code for the modification of the PDE in an OpenFOAM solver for applying an SGS model from our model library is shown in listing~\ref{lst:minSolver}:

\lstset{language=C++,
        basicstyle=\ttfamily,
        keywordstyle=\color{blue}\ttfamily,
        stringstyle=\color{olive}\ttfamily,
        commentstyle=\color{teal}\ttfamily,
        morecomment=[l][\color{magenta}]{\#}
}

\begin{minipage}{\linewidth}
\begin{lstlisting}[caption={A minimal example of the application of the SGS model to a PDE solver in OpenFOAM.}, label=lst:minSolver,basicstyle=\footnotesize]
#include "baseSubGridScaleModel.H"
int main(int argc, char *argv[])
{
    ...
    
    autoPtr<baseSubGridScaleModel> sgsPtr = 
        baseSubGridScaleModel::New(transportProperties.subDict("SGS"));

    while (simple.loop())
    {
        Info<< "Time = " << runTime.timeName() << nl << endl;
        while (simple.correctNonOrthogonal())
        {
            // Compute SGS data needed for scaling.
            sgsPtr->update(psi);
            fvScalarMatrix psiEqn
            (
                fvm::ddt(psi)
              + fvm::div(sgsPtr->scaledFlux(phi, psi), psi)
              - fvm::laplacian(sgsPtr->scaledDiffCoeff(D, psi), psi)
             ==
                fvOptions(psi)
            );

            psiEqn.relax();
            fvOptions.constrain(psiEqn);
            psiEqn.solve();
            fvOptions.correct(psi);
        }
        runTime.write();
    }
}
\end{lstlisting}
\end{minipage}

The SGS requires some values as user input, that are read from the configuration file \mbox{\file{transportProperties}} in OpenFOAM, with the relevant SGS sub-dictionary shown in listing~\ref{lst:transProp}.

\begin{minipage}{\linewidth}
%[language=c++]
\begin{lstlisting}[caption={\file{transportProperties} sub-dictionary for the SGS model}, label=lst:transProp,basicstyle=\footnotesize]
SGS
{
    // Choose the SGS model: "SubGridScale" or "inactive"
    type            SubGridScale;

    // patch to apply SGS to
    patchName       freeSurface;

    // far-field concentration
    cInfinity       cInfinity [0 -3 0 0 1 0 0] 0;

    // (optional:) Switch, whether to limit next
    // search point to be bigger than SMALL
    iSGS            false;

    // (optional:) write parameters for visualization
    // of SGS parameters
    visualizeParameters true;

    // (optional:) verbosity level for info statement output (0-3)
    infoLvl         1;
}
\end{lstlisting}
\end{minipage}

The SGS model can be activated with the keyword-pair \parameter{type SubGridScale} for a mesh boundary patch specified by the keyword \parameter{patchName}. Additional required user input is the far-field concentration, defined with the keyword \parameter{cInfinity}. Optional parameters are, whether to output additional parameters from the SGS model as an OpenFOAM \parameter{volScalarField} and the level of verbosity from 0 to 3 to print to the solver log. 0 corresponds to a quiet mode and 3 prints very extensive information of almost all processes and results within the SGS algorithm and calculations. As these settings affect simulation performance and disk consumption, the first defaults to \parameter{false} while the later affects log readability and defaults to \parameter{1}, which gives limited log-output.
%% Further development
%Another feature of the SGS model library is a rapid info statement level adjustment. Via a label shown in the dictionary entry in listing \ref{lst:transProp}, the output verbosity of the model is controlled in a range from 0 to 3, where 0 is a quiet mode and 3 prints extensive information of almost all processes and results within the SGS calculations.
These visualizing capabilities as well the rapid info statement level enable insights to the user, but they also help with future model development.

\ReviewerTwo{Our framework is continuously and automatically tested, cf \ref{secSub:AutoTesting}.}

\section{Results}
\label{sec:results}

\subsection{\textcolor{Reviewer2}{Error norms and rate of convergence}}
\label{secSub:errNorm_RoC}

\todo[inline]{This whole subsection was added.}

% \subparagraph{Error norms\\}

For verification cases, we employ up to three different error norms.

The $L_1$-, $L_2$- and $L_\infty$-norms are calculated by
% l1norm = np.sum(np.abs(np.subtract(values, analytical_values)))
\begin{align}
    L_1 &= \hspace{13pt} \sum_{i} \left| y_{i,num} - y_{i,ref} \right| \\
    L_2 &= \sqrt{\sum_{i} \left( y_{i,num} - y_{i,ref} \right)^2} \\
    L_\infty &= \hspace{9pt}\max{\left| y_{i,num} - y_{i,ref} \right|}
\end{align}
where $ y_{num} $ is the numerically computed value, $ y_{ref} $ the reference value, either from an analytical solution of the given problem or a value to be assumed correct.
% The $L_2$-norm is calculated by 
% %l2norm = np.sqrt(np.sum(np.power(np.subtract(values, analytical_values), 2)))
% \begin{equation}
%     L_2 = \sqrt{\sum_{i} \left( y_{i,num} - y_{i,ref} \right)^2}
% \end{equation}
% and the $ L_{\infty}$-norm is calculated as
% % LinfNorm = np.max(np.abs(np.subtract(values, analytical_values)))
% \begin{equation}
%     L_\infty = \max{\left| y_{i,num} - y_{i,ref} \right|}.
% \end{equation}

% \subparagraph{Rate of convergence\\}

The rate of convergence is used to describe, how fast a numerical method converges towards a mesh-independent solution.
The rate of convergence for two given cases with different discretization length is approximated as
% c = frac( log(error2/error1) / log(h2/h1) )
\begin{equation}
    c = \frac{\log \frac{e_1}{e_2}}
        {\log \frac{h_1}{h_2}},
\end{equation}
where $c$ is the rate of convergence, $e$ is the error, computed as $e = y_{num} - y_{ref}$, and $h$ is the typical discretisation length.
For the cases, where we compute a rate of convergence in this article, the edge length of the surface mesh is the typical discretization length.
The subscripts ${}_1$ and ${}_2$ denote the two different cases, where $h_1$ is the bigger discretisation length.

\newcommand{\testFP}{flat plate }

\subsection{Flat plate test case}
\label{secSub:test_flatPlate}

\begin{figure}[!htbp]
    \centering
    \includegraphics[width=0.9\textwidth]{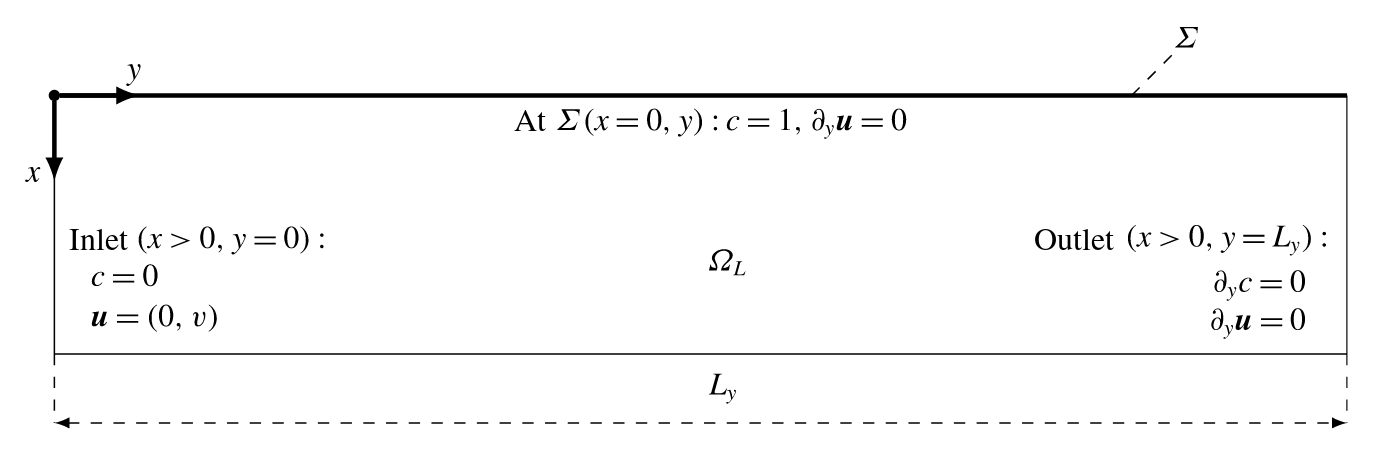}
    \caption{Set-up of the flat plate test. Figure taken from \cite{pesci_SGS_risBubb_surfactants_2018}.}
    \label{fig:flatPlate_setup}
\end{figure}

The smallest test for model development and code verification is what we call the \testFP test case. It implements the assumptions made in \cref{secSub:SGS_anaModel} and was already implemented by e.g.~\cite{pesci_SGS_risBubb_surfactants_2018}.
The test consists of a flat plate with a prescribed constant concentration of $ \SI{1}{mol/m^3} $.
%Parallel to this plate a flow is prescribed, that has no interaction with the wall, i.e.\ there is no velocity boundary layer or velocity gradient.
\ReviewerTwo{A velocity parallel to the plate is prescribed for the whole domain, including the boundaries so there is no velocity boundary layer at the plate.}
Initially and in the inflow, the concentration is set to zero, the length of the plate is $\SI{0.005}{m}$ and the fluid velocity is $\SI{0.1}{m/s}$. The timestep of $\SI{0.0002}{s}$ is used for a physical duration of $\SI{0.1}{s}$, which is double the time the fluid needs to be transported through the domain.
\Cref{fig:flatPlate_setup} visualizes the setting.

%Depending on the grid size, the velocity and the diffusivity, different behaviour emerges.
With the possibilities described in \cref{text:visualizeFigs}, it can be verified that fundamental parts of the algorithm as described in \cref{secSub:SGS_algo} are implemented correctly, like the boundary layer thickness~$ \delta $ and the diffusion correction. %The convection correction is not relevant to this test case since the velocity is parallel to the boundary.
Beneficially, there is an analytical solution to this model for comparison.
For these reasons the test enables insights into the capabilities of the SGS model and the implementation thereof. This simple test case does circumvent most of the error-handling and also convective flux correction must not have an effect on the outcome.

\begin{figure}[!htbp]
    \centering
    \includegraphics[width=\textwidth]{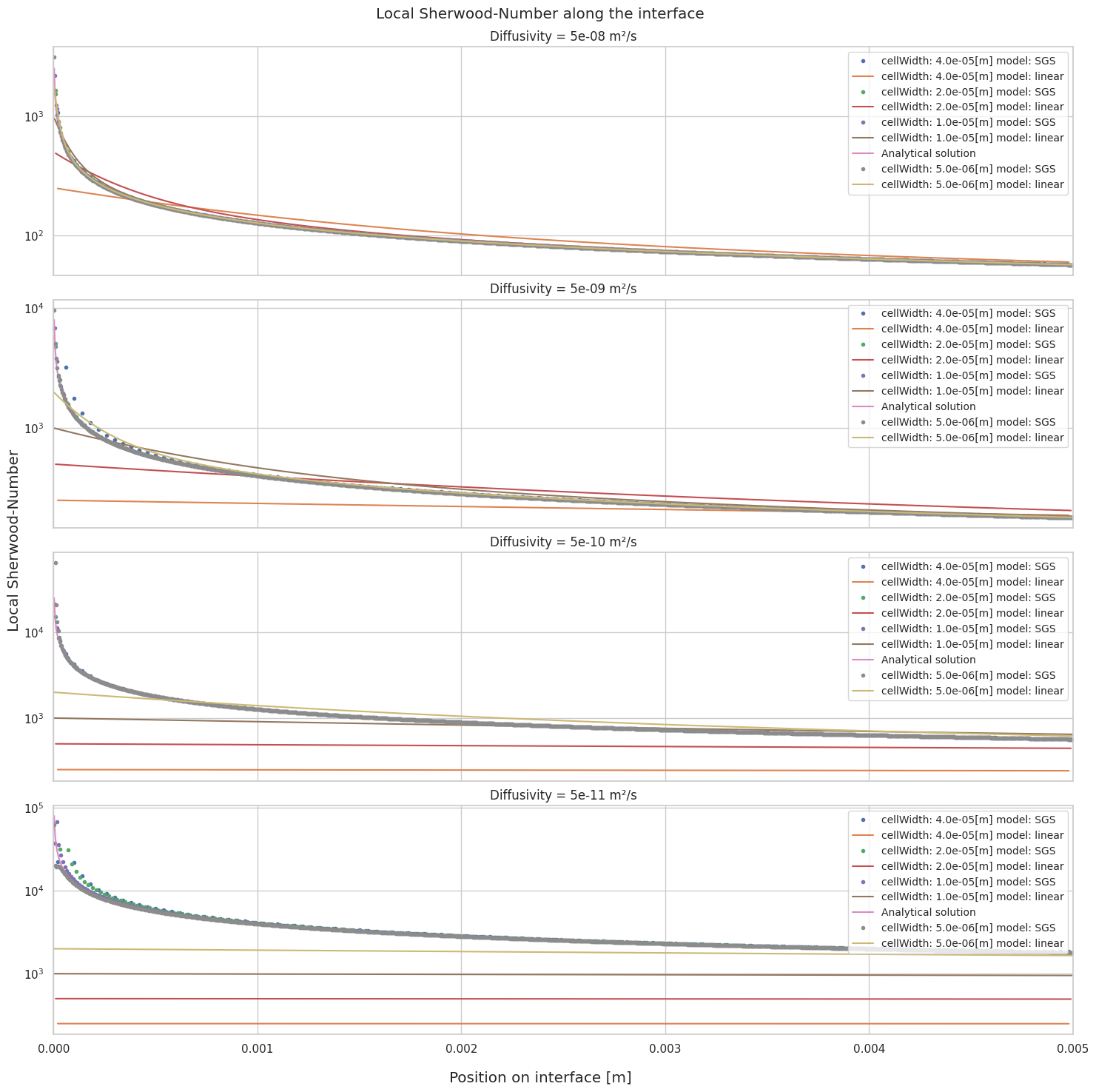}
    \caption{Results for the parameter study over of the flat plate test case}
    \label{fig:flatPlate_results}
\end{figure}
In \cref{fig:flatPlate_results} the local Sherwood numbers of a parameter study along the boundary surface are displayed. The parameter study has four levels of mesh refinement, where the cell size halves between each execution from $ \SIrange{40}{5}{\mu m} $, four different diffusivity levels spanning four orders of magnitude from $\SIrange{5e-8}{5e-11}{m^2/s}$ and every of these setups is run with and without active SGS model. Taking the length of the plate as reference length $ L $, the Peclet number is within the interval $[10^4, 10^7]$.

It is apparent, how without the SGS model, the Sherwood number cannot be computed in cases with a boundary layer that is small in comparison to the cell size, i.e.\ when the cell size is big and the diffusivity is small. In contrast to the linear interpolation with the SGS model the Sherwood number is captured reasonably well in all cases.
Note in \cref{fig:flatPlate_results}, that at some locations the Sherwood number with the application of the SGS model is actually smaller than without it. 

\subsection{\otherChange{Mass transfer at a spherical fluid particle}}
\label{secSub:test_sgsWedge}

The second test case for the SGS model mimics a moving droplet or bubble. It uses the axisymmetric solution of velocity field from Satapathy and Smith in \cite{satapathy_motion_1961} for low Reynolds numbers. The computational domain is a wedge with prescribed velocity. This test case was also implemented by and we take the reference solution from \cite{pesci_SGS_risBubb_surfactants_2018}.
This test case introduces additional complexity compared to \cref{secSub:test_flatPlate}, since there is flow non-parallel to the interface, therefore violating some of the assumptions made when deriving the SGS mathematical model, as described in \cref{secSub:SGS_anaModel}. From the visualization capabilities we can see in the rear part of the "bubble" the exception handling partially taking place, as there the concentration rises, not conforming to the assumption of a thin boundary layer (with parallel flow) made in the derivation of the SGS.
As in \cref{secSub:test_flatPlate} the inflow and initial concentration is set to $\SI{0}{mol/m^3}$ and the interface is set to have a constant concentration of $\SI{1}{mol/m^3}$. The simulated physical duration is $\SI{0.6}{s}$ and the results shown here represent the final state.

To see the best results with the SGS, it is important to set the interpolation schemes to \parameter{limitedLinear phi 1.0} for the concentration c and to \parameter{linear} for the interpolation of the scaled diffusion coefficient as well as for the flux of U in the \file{fvSchemes}-file.

\begin{figure}[!htbp]
    \centering
    \includegraphics[width=0.9\textwidth]{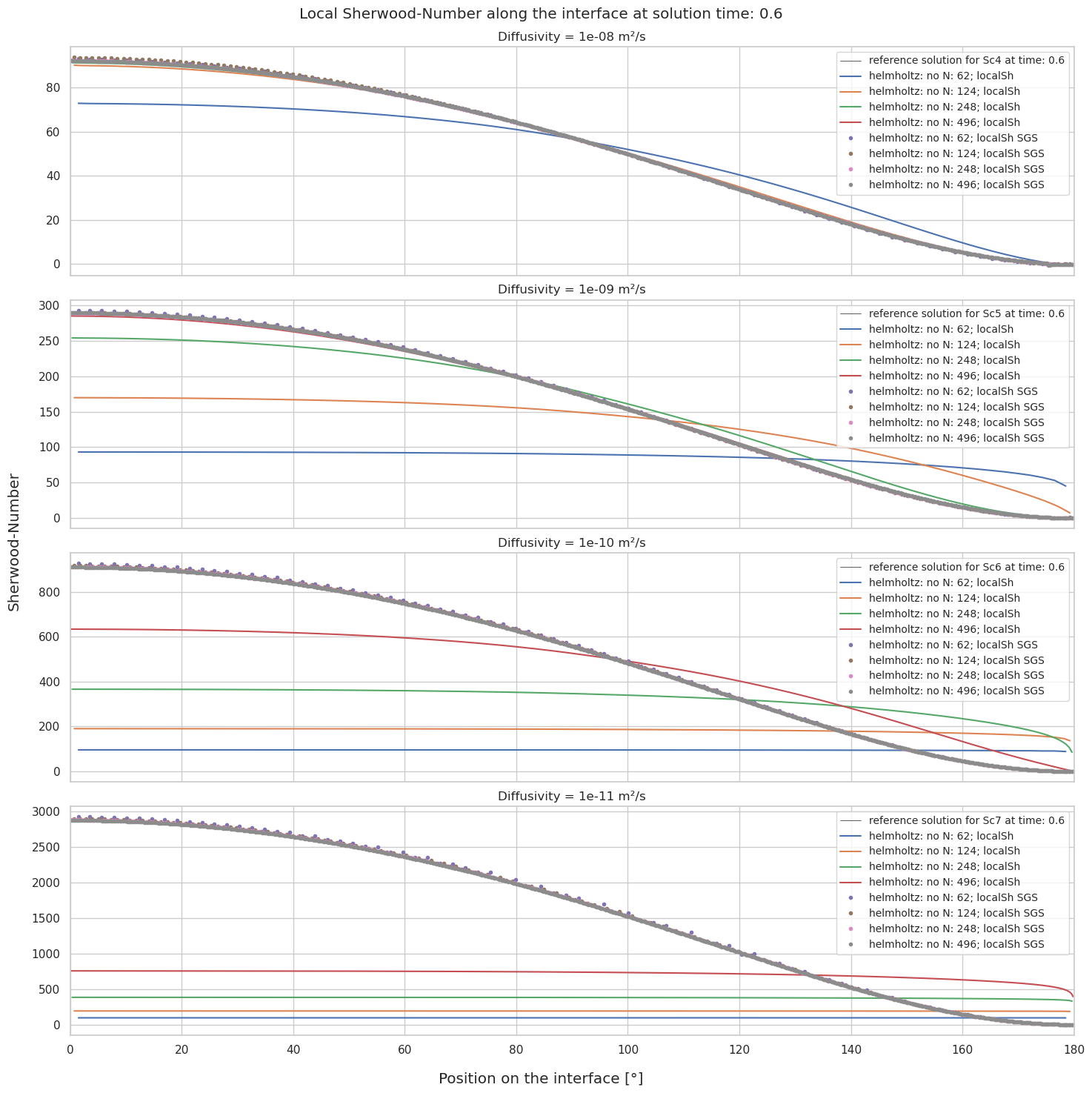}
    \caption{Local Sherwood numbers of a parameter study for Schmidt-numbers from $\SIrange{1e4}{1e7}{}$}
    \label{fig:sgsWedge_results}
\end{figure}

\begin{sloppypar}
We run a parameter study consisting of four different diffusivities from ${ \SIrange{1e-8}{1e-11}{m^2/s} }$, four different mesh resolutions with the meshing parameter $\rm N = [62, 124, 248, 496] $ as the number of faces along the half-circled interface. Every study is run with and without employing the SGS and with a constant timestep of $\SI{0.0002}{s}$. The bubble diameter of $\SI{2}{mm}$ serves as reference length for the Sherwood number.
It can be observed from \cref{fig:sgsWedge_results} that without deploying the SGS, the local Sherwood number is almost constant and except for the rear part of the bubble too low in case of small diffusivities and coarse meshes. When the mesh resolution is fine and the diffusivity is high, the boundary layer is not as thin (in comparison to the mesh cell size), and therefore the results of linear modeling and SGS modeling are similar.
\end{sloppypar}

\begin{figure}[!htbp]
    \centering
    \includegraphics[width=0.9\textwidth]{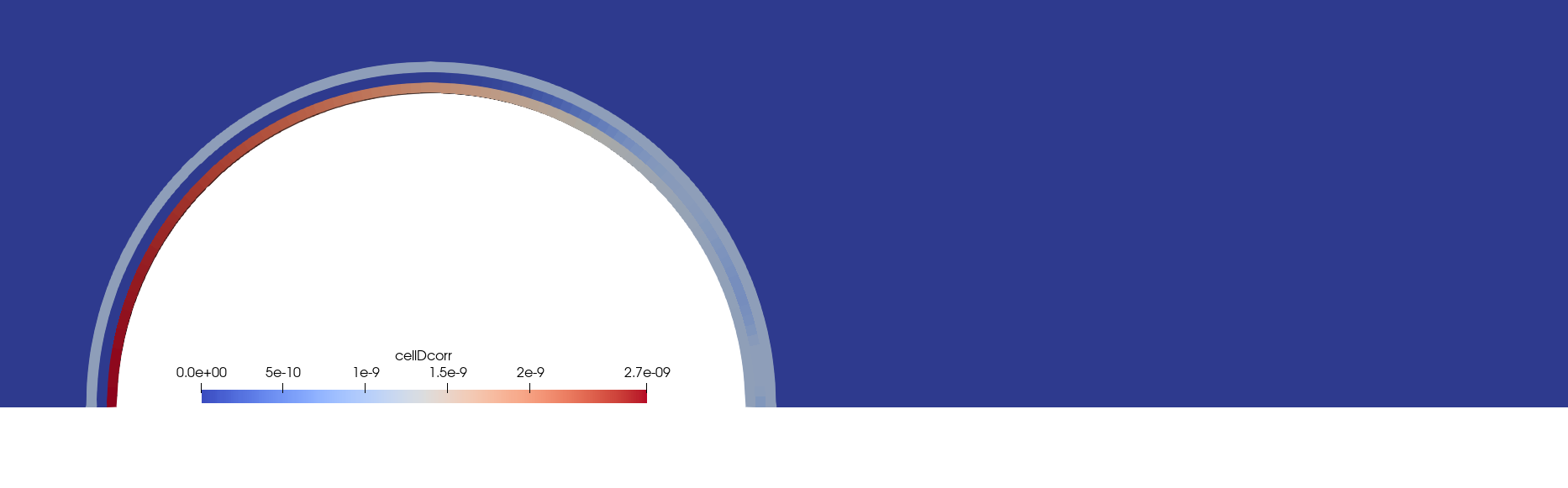}
    \caption{Visualization of corrected diffusion coefficients for the coarsest mesh with a diffusivity of $\SI{1e-9}{m^2/s}$}
    \label{fig:sgsWedge_Dcorr}
\end{figure}

In \cref{fig:sgsWedge_Dcorr} the diffusion coefficient after scaling with the SGS is displayed. In the cell layer adjacent to the interface, the diffusion coefficient at the interface is visible, whilst in the next outer cell layer the scaled diffusion coefficient at the faces between interface-adjacent cells and that same cell is visualized. The third row of cells represents the uncorrected diffusion coefficient for reference.

The biggest correction is applied in the front part of the bubble, where the flow impinges onto the interface and the boundary layer is very thin, therefore the derivative in surface normal direction of the concentration is very big. % surface normal gradient is very steep.
In the rear part of the bubble the boundary layer has grown and the main direction of the flow is no longer parallel to the surface, making the surface concentration gradient in interface-normal direction much smaller and reasonable good represented by linear interpolation. Therefore less correction is needed. To understand what is happening, we would like to point the reader to \cref{fig:sgs_concProfile} again.

\subsection{Marangoni\otherChange{-induced} migration of an air bubble in silicone oil}
\label{secSub:results-marangoni}

The velocity of a bubble/droplet within an unbounded fluid subjected to uniform temperature gradient ($\partial T/\partial z$) and zero gravity field ($g=0$) is defined by following expression \cite{Young1959}:
\begin{equation}
    v^T_{\rm b, YBG} = 
    \frac{2}
    {
        \left(2+\dfrac{\lambda_i}{\lambda_o}\right)
        \left(2 + 3\dfrac{\mu_i}{\mu_o}\right)
    }
    \frac{R}{\mu_o}
    \frac{\partial \sigma}{\partial T}
    \frac{\partial T}{\partial z},
    \label{eq:BubbleMigrationVelocity}
\end{equation}
where $R$ is the radius of the bubble, $T$ is the fluid temperature, $\sigma$ is the surface tension coefficient, $\mu_i$ is the dynamic viscosity of the internal fluid, $\mu_o$ is the dynamic viscosity of the outside fluid and $\lambda$ is the thermal conductivity of the fluid.

In the case the Marangoni convection is caused by a specified fixed gradient of surfactant concentration along the interface, the bubble migration velocity can be expressed based on \cref{eq:BubbleMigrationVelocity} as follows, see \cite{Muradoglu2008}:
\begin{equation}
    v^\Gamma_{\rm b, YBG} = 
    \frac{2}
    {
        3
        \left(2 + 3\dfrac{\mu_i}{\mu_o}\right)
    }
    \frac{R}{\mu_o}
    \frac{\partial \sigma}{\partial \Gamma}
    \frac{\partial \Gamma}{\partial z},
    \label{eq:BubbleMigrationVelocity2}
\end{equation}
where $\Gamma$ is the concentration of surfactants at the interface.

A numerical simulation is performed for an air bubble of radius $r_b = 1~{\rm mm}$ moving through the silicone oil. Relevant properties of both phases are listed in \cref{tab:properties}. The bubble is assumed to be of spherical shape and rigid and is fixed during the simulation. The spatial computational domain consist of a volume of the spherical bubble filled with air and volume around the bubble bounded by the spherical surface of radius $r=20r_{\rm b}$ and filled with silicone oil. The spatial domain is discretised by the unstructured mesh consisting of $\num{979520}$ hexahedral cells.

The flow inside and outside of the bubble is driven by Marangoni forces due to the nonuniform distribution of surfactant concentration along the interface, which is fixed during the simulation. The distribution of surfactant along the bubble surface is defined by following linear function:
\begin{equation}
    \frac{\Gamma}{\Gamma_\infty} = \frac{z + 2r_{\rm b}}{L},
\end{equation}
where $\Gamma_\infty$ is the saturated surfactant concentration and $L$ is the length scale, whose value is set to $\SI{525}{mm}$. The equation of state is also linear and reads as follows:
\begin{equation}
    \sigma = \sigma_{\rm s}\left(1 - \beta_{\rm s}\frac{\Gamma}{\Gamma_\infty}\right),
\end{equation}
where $\beta_{\rm s} = \frac{\mathcal{R}T\Gamma_\infty}{\sigma_{\rm s}}$ is the elasticity number which amounts to $2$ in this study.

The unsteady computation is performed with time-step size $\Delta t = \SI{0.01}{s}$ using the first order accurate \parameter{Euler} temporal discretisation scheme, until steady state is reached. The gradient of the surface tension coefficient as well as the gradients of volume fields is calculated using the least square method. The diffusion terms in the momentum and pressure equations are discretised using the skew-corrected central differencing scheme, while the convective term in the momentum equation is discretised using the linear-upwind scheme. At the outer side of the spatial domain the velocity and normal derivative of pressure are set to zero and at the interface corresponding kinematic and dynamics conditions are enforced.

\begin{figure}[!htbp]
    \centering
    \includegraphics[width=1\linewidth]{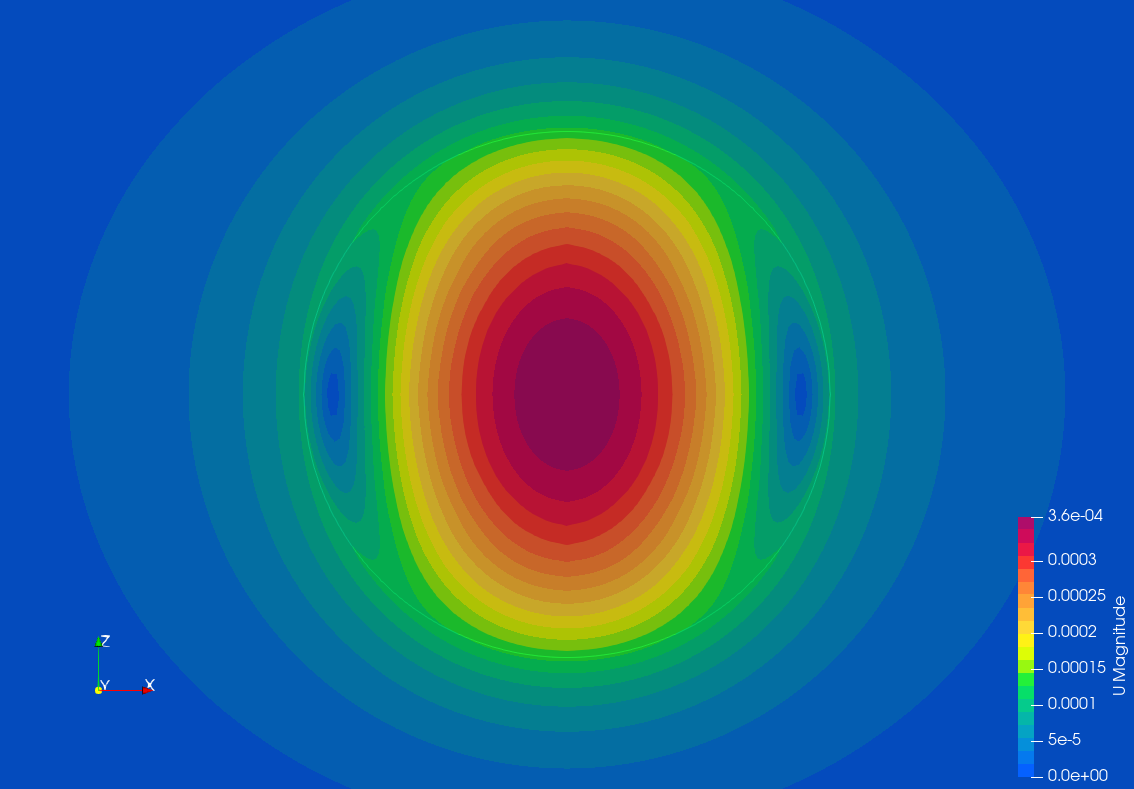}
    \caption{\otherChange{Marangoni-induced velocity magnitude} inside and around the \otherChange{migrating} bubble along the plane $y=0$}
    \label{fig:marangoni-bubble-velocity}
\end{figure}

The bubble rise velocity is calculated based on numerically determined volume flow rates through the faces at the interface as follows:
\begin{equation}
    v_{\rm b} = \frac{\sum_f {\rm pos}(\phi_{\rm i,f})\phi_{\rm i,f}}{\sum_f {\rm pos}(\vec{k}\cdot\vec{S}_{\rm i,f}) (\vec{k}\cdot\vec{S}_{\rm i,f}) },
    \label{eq:bubble-velocity}
\end{equation}
where $\phi_{\rm i,f}$ is the volume flow rate of the fluid through the face $f$ at the interface, $\vec{S}_{\rm i,f}$ is the area vector of the face $f$ at the interface and $\vec{k}$ is the unit direction vector of bubble motion. The bubble velocity calculated using \cref{eq:bubble-velocity} based on simulation results amounts to $v_b = \SI{0.001325}{m/s}$, while the corresponding analytical (\cref{eq:BubbleMigrationVelocity2}) gives the migration velocity as $v^\Gamma_{\rm b,YBG}= \SI{0.0013295}{m/s}$. The velocity field inside and around the bubble along the plane $y=0$ is shown in \cref{fig:marangoni-bubble-velocity}.

\begin{table}
    \centering
    \setlength{\tabcolsep}{8pt} % Adjust column separation
    \renewcommand{\arraystretch}{1.3} % Adjust row height
    \begin{tabular}{|c|c|c|c|}
    \hline
         & $\rho$ [kg/m$^3$] & $\mu$ [Pa s]  & $\sigma$ [N/m]\\
    \hline
        Air & $1.8\times 10^{-5}$ & $1.2$ & $0.02$\\
    \hline
        Oil & $955$ & $0.191$ & $0.02$\\
    \hline
    \end{tabular}
    \caption{Properties of air and silicone oil}
    \label{tab:properties}
\end{table}

\subsection{\otherChange{Passive scalar transport on the interface}}
\label{secSub:rotatingDroplet}

We test the framework's ability of adequate transport of a species along an interface with a rotating droplet with a prescribed concentration. The droplet is represented in 2-D. % also called cylinder
The test setup follows the test done by \cite{AntritterDiss2022, antritter_two-field_2024}. We prescribe the initial concentration distribution and the velocity. For the evaluation we compute one rotation, what equals one second with our prescribed velocity referring to an angular velocity $ \omega = 1/s $.

The initial distribution is given by
\begin{equation} \label{eq:rotating_iniDistr}
    \Gamma (\theta, \tau = 0) = \frac{\cos{\theta} + 1}{2} \Gamma_0 ,
\end{equation}
where $ \Gamma $ is the surface concentration, $ \tau $ is the time, $ \theta $ is the angle and $ \Gamma_0 $ is the maximum initial concentration.
After a time $ \tau $ the expected distribution from an analytical approach is given by
\begin{equation}
    \Gamma (\theta, \tau) = \frac{ e^{-D_I \tau} \cos( \theta - \omega \tau ) + 1}{2} \Gamma_0 ,
    %\label{eq:rotating_finalDistr}
\end{equation}
where $ D_I $ is the diffusion in interface tangential direction.

\begin{figure}[!htbp]
    \centering
    \includegraphics[width=0.9\textwidth]{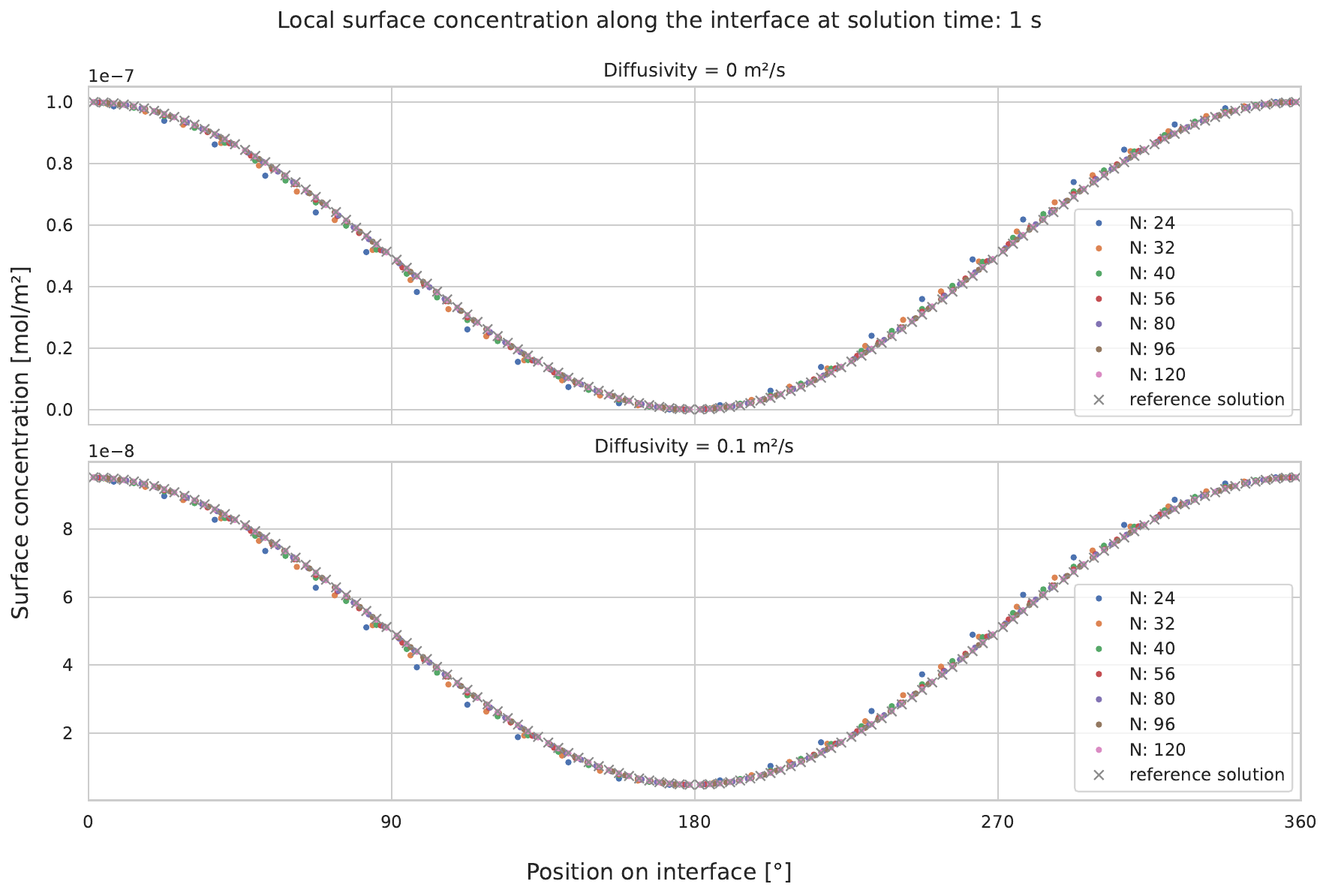}
    \caption{Results from the rotating droplet parameter study}
    \label{fig:rotating_ParamStudy_results}
\end{figure}

\begin{sloppypar}
For proper evaluation a mesh refinement study with a parameter variation of the diffusivity is conducted. The results are visualized in \cref{fig:rotating_ParamStudy_results}. 
\ReviewerTwo{We use a meshing parameter $N = [24, 32, 40, 56, 80, 96, 120] $, corresponding to the number of surface cells along the interface.}
The timestep is set to $\SI{0.0005}{s}$ regardless of the mesh size. The surface diffusivity is varied between $\SIrange[range-phrase=\,\rm and\,]{0}{0.1}{m^2/s}$.
\end{sloppypar}

% https://tex.stackexchange.com/questions/37581/latex-figures-side-by-side
% figures side by side: 1 figure with 2 labels/subfigures or 2 "real figures"
\begin{figure}[!htbp]
    %\centering
    \begin{subfigure}{.5\textwidth}
        \centering
        \includegraphics[width=0.95\linewidth]{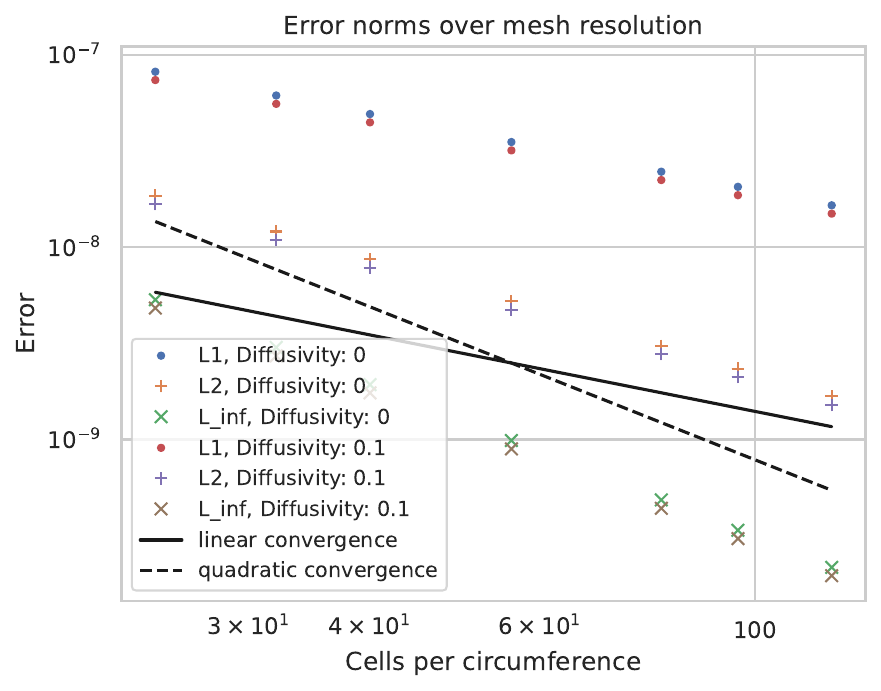}
        \caption{Error norms}
        \label{fig:rotating_ParamStudy_errorNorms}
    \end{subfigure}
    \begin{subfigure}{.5\textwidth}
        \centering
        \includegraphics[width=0.95\linewidth]{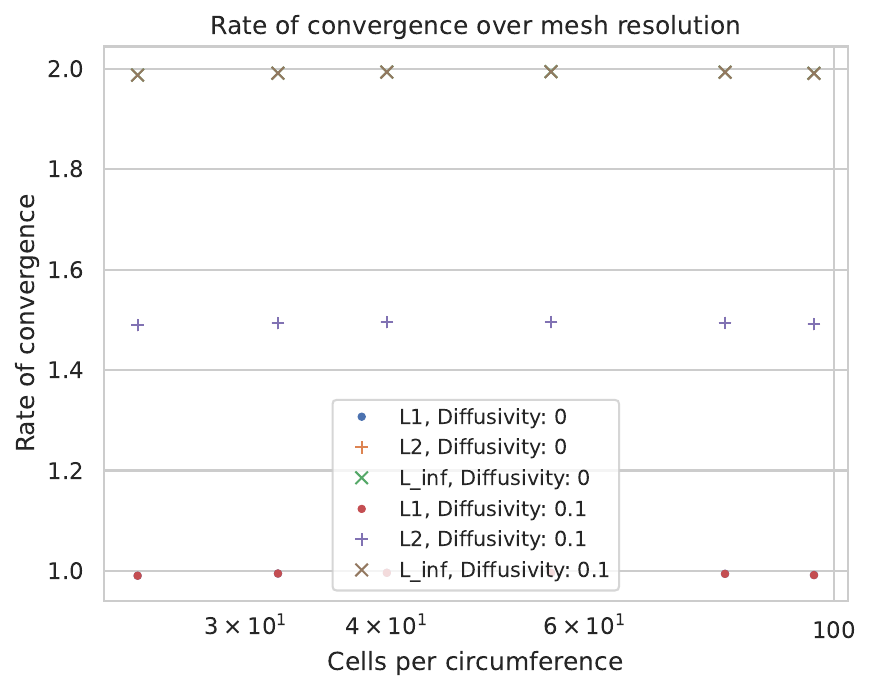}
        \caption{Rate of convergences}
        \label{fig:rotating_ParamStudy_RoC}
    \end{subfigure}
    \caption{Errors of the rotating droplet parameter study}
    \label{fig:rotating_ParamStudy_errorsMainFig}
\end{figure}

\begin{comment}
\begin{figure}[!htbp]
    \centering
    \includegraphics[width=0.6\textwidth]{figures/results/rotatingContaminatedDroplet/output_rotating_errorNorms.png}
    \caption{Error norms of the rotating droplet parameter study}
    \label{fig:rotating_ParamStudy_errorNorms}
\end{figure}

\begin{figure}[!htbp]
    \centering
    \includegraphics[width=0.6\textwidth]{figures/results/rotatingContaminatedDroplet/output_rotating_RateOfConvergences.png}
    \caption{Rate of convergences of the rotating droplet parameter study}
    \label{fig:rotating_ParamStudy_RoC}
\end{figure}
\end{comment}

%For coarse meshes we see a strong shift in the results, as if the droplet would turn. Already by the results, we see clear convergence with finer mesh resolution. In \cref{fig:rotating_ParamStudy_errorNorms} we can see, that the case converges very good and that the cases with a little bit of diffusion have slightly lower error norms in all cases.
%(TODO: keep (1st half)sentence? It seems the diffusivity stabilizes the solution and actually would allow for much larger time steps.)

Visually it is difficult to see a difference between the discretizations in \cref{fig:rotating_ParamStudy_results},
\ReviewerTwo{except for very coarse mesh resolutions}.
In \cref{fig:rotating_ParamStudy_errorsMainFig}
%\cref{fig:rotating_ParamStudy_errorNorms,fig:rotating_ParamStudy_RoC} % no space after comma!
the error norms and their rate of convergence are shown.
\ReviewerTwo{The rate of convergence of the $L_\infty$-norm is close to~$2$, for the $L_2$-norm it is close to~$1.5$ and for the $L_1$-norm is around $1$ for the resolutions analyzed.
The rates of convergence shown in \cref{fig:rotating_ParamStudy_RoC} are computed from the mesh resolution, at which they are shown and the next finer one.}
With the low error norms and their respective rate of convergence we deduct, that our framework is well able to handle transport along an interface.

\subsection{Expanding fluid interface}
\label{secSub:expandingDroplet}

To test the frameworks conservation of surfactant mass for an expanding \otherChange{fluid interface}, we implement a test case, that mimics an expanding droplet like it was done by~\cite{AntritterDiss2022,antritter_two-field_2024}. The initial distribution of species is described and there is no diffusion happening. The initial distribution is the same as in \cref{eq:rotating_iniDistr}. The droplet is represented by a wedge that is inflated with a prescribed velocity and thereby the surface is stretched. The analytical solution to this setup is given by
\begin{equation}
    \Gamma (\theta, r(\tau)) = \frac{ \cos{\theta} + 1}{2} \Gamma_0 \frac{r_0^2}{r(\tau)^2},
\end{equation}
where $ r(\tau) = \sqrt[3]{3 V_0(\tau + \tau_0)} $, where $ \tau_0 = 1/3 s $ and $ r $ is the radius, respectively $ r_0 $ is the initial radius.

\begin{figure}[!htbp]
    \centering
    \includegraphics[width=0.5\textwidth]{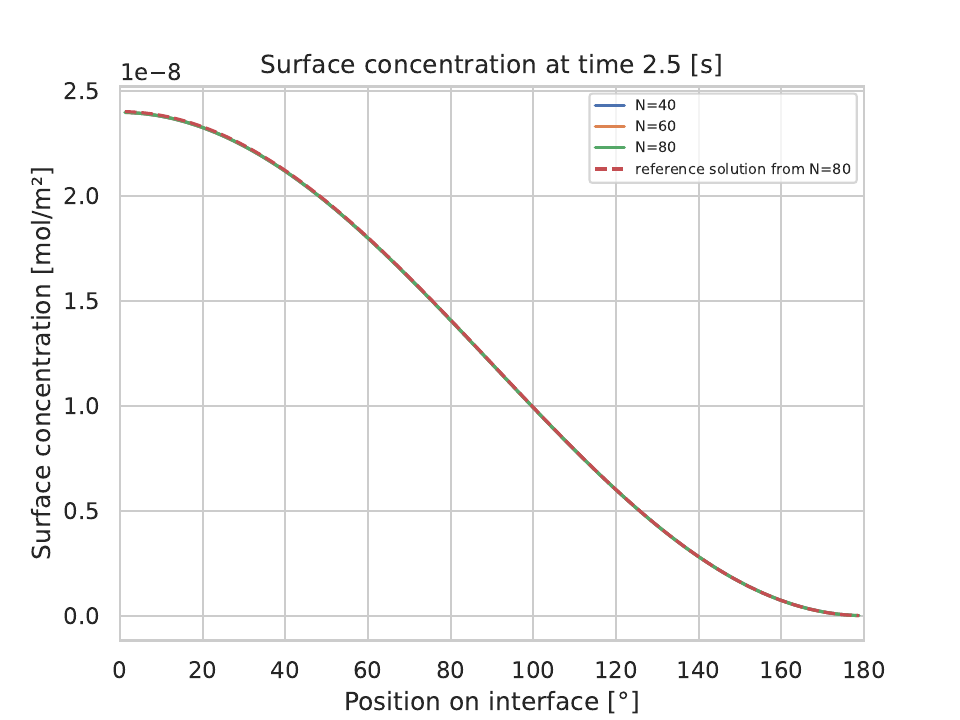}
    \caption{Surface concentration after $\SI{0.5}{s}$ for different mesh resolutions}
    \label{fig:expanding_CsOverTime}
\end{figure}

\Cref{fig:expanding_CsOverTime} shows, how the species concentration per area at the interface has decreased after $\SI{2.5}{s}$ of bubble growth.
% Notably there is some deviation from the expected results near the axisymmetry line at the angular position of 0°. NOT ANYMORE
The concentration matches precisely the expected result for all mesh resolutions tested. We conclude, that our framework is fit to simulate phenomena involving surface expansion/shrinking.

\subsection{\ReviewerTwo{Bubble rising in surfactant solution}}
\label{secsub:rising_bubble}

\todo[inline]{This whole case was added.}

To showcase the framework's capabilities, we add a complex test case.
This test case is a re-calculation of the case from \citep{pesci_SGS_risBubb_surfactants_2018}.
It models a single air bubble rising in quiescent water with a surfactant, and there is experimental data to validate our simulation.

The previous test cases have shown in an isolated way, that the framework is able to resolve the boundary layer (\cref{secSub:test_flatPlate,secSub:test_sgsWedge}), handle chemically induced changes in surface tension (\cref{secSub:results-marangoni}), compute surfactant transfer on the interface (\cref{secSub:rotatingDroplet}) and conserve surfactant mass on the interface, when the interface area changes (\cref{secSub:expandingDroplet}).
For this test case all of these capabilities are needed.

\begin{table}[!htbp]
    \centering
    \setlength{\tabcolsep}{8pt} % Adjust column separation
    \renewcommand{\arraystretch}{1.3} % Adjust row height
    \begin{tabular}{|c|c|c|c|c|c|}
        \hline
        $\rho^+$ [kg/m$^3$] & $\mu^+$ [kg/ms] & $\rho^-$ [kg/m$^3$] & $\mu^-$ [kg/ms] & $\sigma_0$ [N/m] \\ \hline
        997.3 & 9.3 $\times 10^{-4}$ & 1.1965 & 1.83 $\times 10^{-5}$ & 0.0724 \\
        \hline
    \end{tabular}
    \caption{Fluid properties to model water (${}^{+}$) and air (${}^{-}$)}
    \label{tab:fluid_properties}
\end{table}

\begin{figure}[!htbp]
    \centering
    \includegraphics[width=1.0\textwidth]{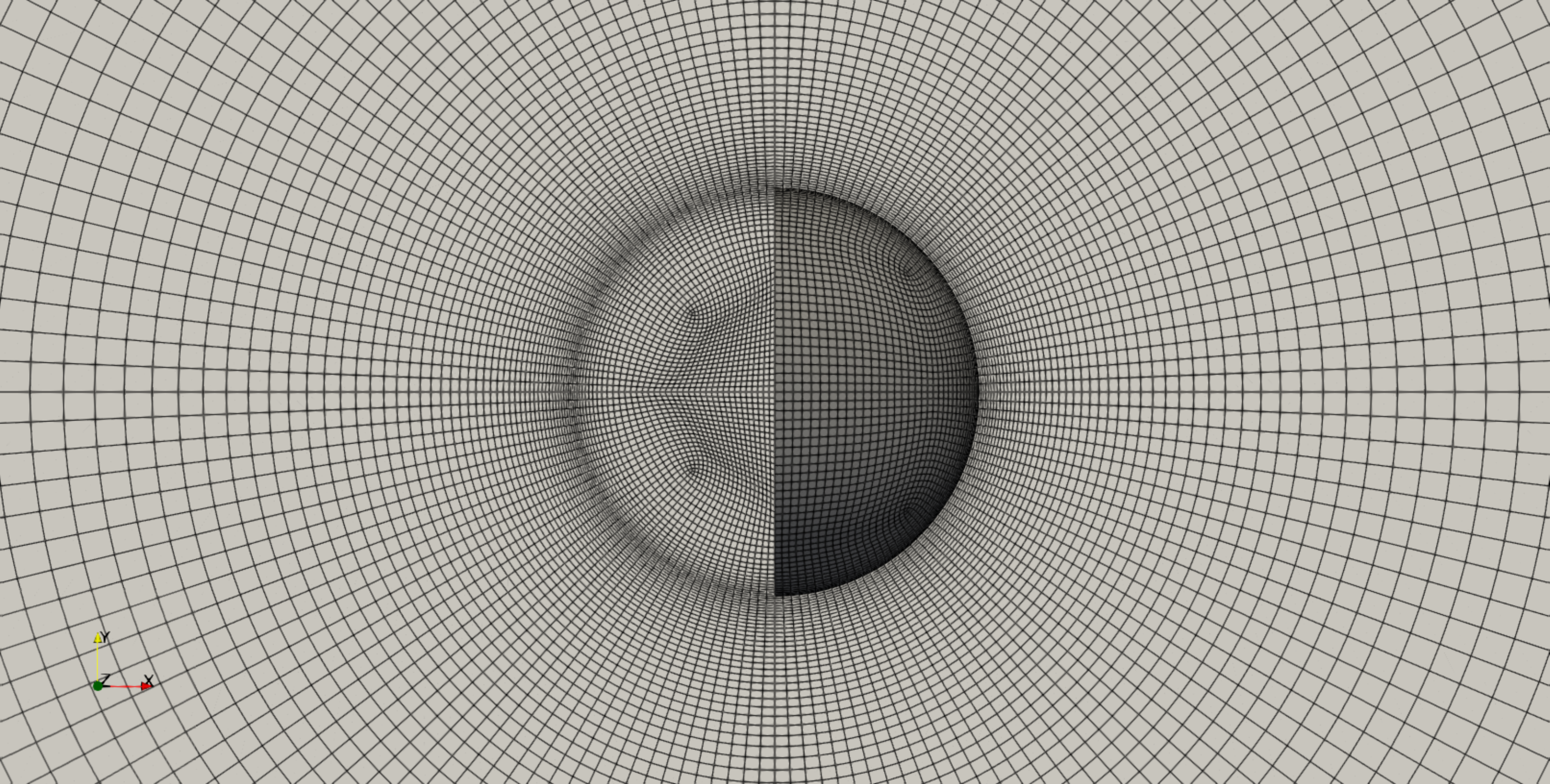}
    \caption{Undeformed mesh and bubble at the beginning of the simulation}
    \label{fig:risBubble_mesh}
\end{figure}

As we will compare our results with \citep{pesci_SGS_risBubb_surfactants_2018}, we strive to use exactly the same settings in \citep{pesci_SGS_risBubb_surfactants_2018}. However, \citep{pesci_SGS_risBubb_surfactants_2018} have used an ALE-IT implementation in another OpenFOAM version, which introduces some  differences in the solution algorithm, addressed below.
The bubble diameter is \SI{1.45}{mm}, and it is initialized as a sphere with zero velocity.
The fluid properties used can be found in \cref{tab:fluid_properties}.
The timestep size used is \SI{5e-6}{s} and the mesh contains \num{979520}~cells, where the interface is being discretized with \num{9280}~faces.
It is displayed in \cref{fig:risBubble_mesh}, where on the right-hand side, the interface mesh is displayed in dark grey.
The mesh is created with OpenFOAM's \parameter{blockMesh} utility and, therefore, is different than the mesh in \citep{pesci_SGS_risBubb_surfactants_2018}, but it has similar discretization lengths.
Also, the newly introduced switch to limit the boundary layer thickness to a minimum of \SI{1e-15}{m} is de-activated to perform a more direct comparison with \citep{pesci_SGS_risBubb_surfactants_2018}.

To track the bubble's motion a moving reference frame (MRF) is used.
This MRF follows the bubble, so it stays in the center of the computational domain and the mesh is less deformed.
The domain is now moving and accelerating like the bubble, making a correction in the momentum equation (adding $\rho \alpha_{MRF}$ as well as a modified boundary condition for the velocity ($ v_{out} = - v_{MRF} $) necessary.

The surfactant modelled is the non-ionic dodecyl-dimethyl-phosphine-oxide~($C_{12}DMPO$).
Its properties can be found in \cref{tab:C12DMPO_properties}.
The sorption process is modelled with the fast Langmuir model and the thin concentration boundary layer at the interface is treated with the SGS model.

\begin{table}[!htbp]
    \centering
    \setlength{\tabcolsep}{8pt} % Adjust column separation
    \renewcommand{\arraystretch}{1.3} % Adjust row height
    \begin{tabular}{|c|c|c|c|c|}
        \hline
        $c_\infty^\Sigma$ [mol/m$^2$] & $a_L$ [mol/m$^3$] & $D$ [m$^2$/s] & $D^\Sigma$ [m$^2$/s] & $T$ [K] \\ \hline
        4.17 $\times 10^{-6}$ & 4.85 $\times 10^{-3}$ & 5 $\times 10^{-10}$ & 5 $\times 10^{-7}$ & 296 \\
        \hline
    \end{tabular}
    \caption{Surfactant (C$_{12}DMPO$) properties, including fast Langmuir adsorption model parameters}
    \label{tab:C12DMPO_properties}
\end{table}

We model 4 different cases, where each one starts with a clean interface.
The liquid phase is initialized with a surfactant concentration of $ c_{initial} = c_\infty = [0,\ \num{2e-3},\ \num{8e-3},\ \num{5e-2}]\ \si{mol / m^3}$, where $c_\infty $ is also the concentration boundary condition for inflow into the liquid domain.
There is no surfactant present in the gas phase at any point.

\begin{figure}[!htbp]
    \centering
    \includegraphics[width=0.75\textwidth]{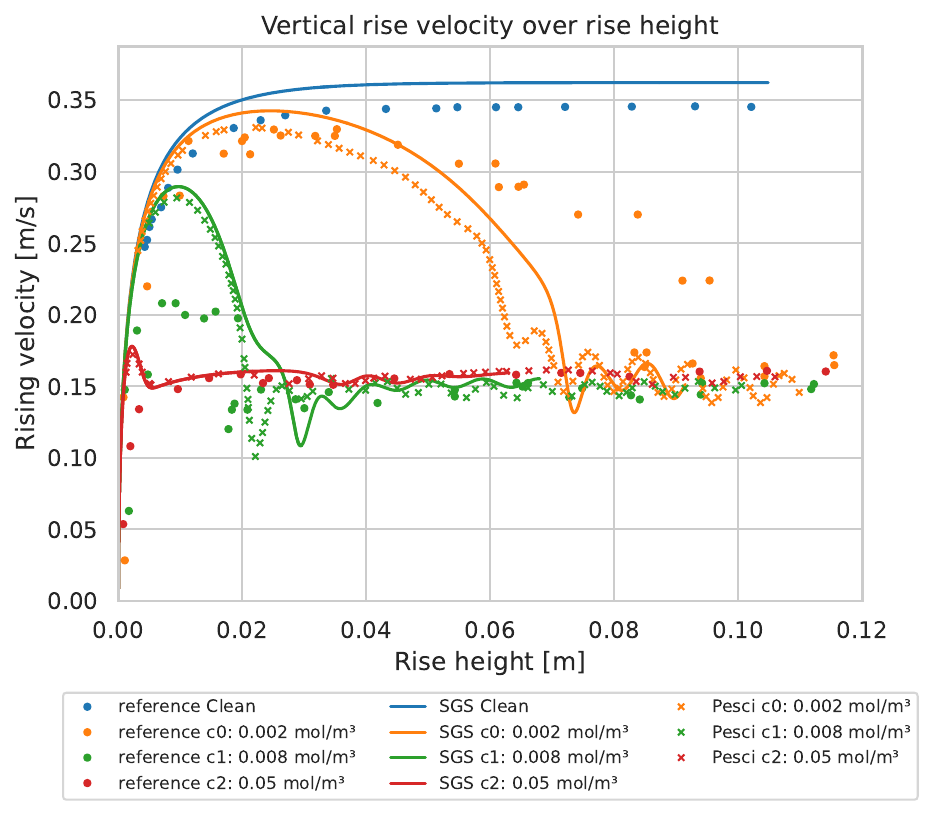}
    \caption{Rising velocities over rising height for different initial concentrations}
    \label{fig:risBubble_velos}
\end{figure}

\Cref{fig:risBubble_velos} shows the rising velocities obtained by experiments \citep{pesci_experimental_2017_book}, previous simulations  \citep{pesci_SGS_risBubb_surfactants_2018}, as well as results of our new ALE-IT implementation.
There are only minor discrepancies between the two solutions obtained by ALE-IT simulations with SGS, that are far exceeded by the differences to the experimental results.
Small discrepancies between two different ALE-IT methods are caused by the differences in the numerical setup, as well as uncertainties related especially to the surfactant properties, like the surface diffusion coefficient ,different meshing, slightly different numerical schemes, and the Finite Element solver used for the mesh motion in \citep{pesci_SGS_risBubb_surfactants_2018}.
Both sets of results show the characteristic features of over- and undershooting for all initial concentration levels and a good overall agreement with experimental data.

% \begin{figure}[!htbp]
%     \centering
%     \includegraphics[width=0.8\textwidth]{figures/results/risingBubble/2e-3/surfactant_0-109_fieldAndSurface.png}
%     \caption{PLACEHOLDER IMAGE!!!: Surfactants on the surface and in the bulk phases}
%     \label{fig:risBubble_surfactDistr}
% \end{figure}

% https://tex.stackexchange.com/questions/37581/latex-figures-side-by-side
% figures side by side: 1 figure with 2 labels/subfigures or 2 "real figures"
\begin{figure}[!htbp]
    \centering
    \begin{subfigure}{.3\textwidth}
        \centering
        \includegraphics[width=0.95\linewidth]{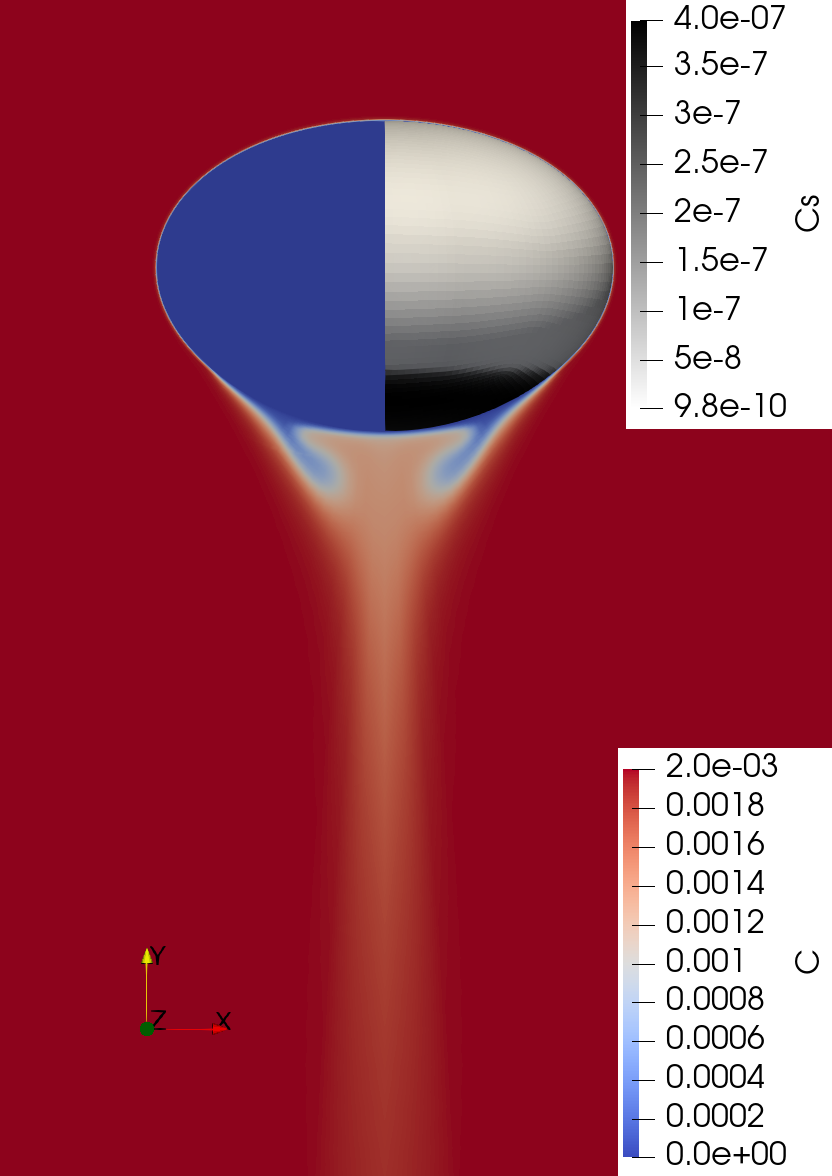}
        \caption{$c_{\infty} = $ \SI{2e-3}{mol/m^3}}
        % \label{fig:rotating_ParamStudy_errorNorms}
    \end{subfigure}
    \begin{subfigure}{.3\textwidth}
        \centering
        \includegraphics[width=0.95\linewidth]{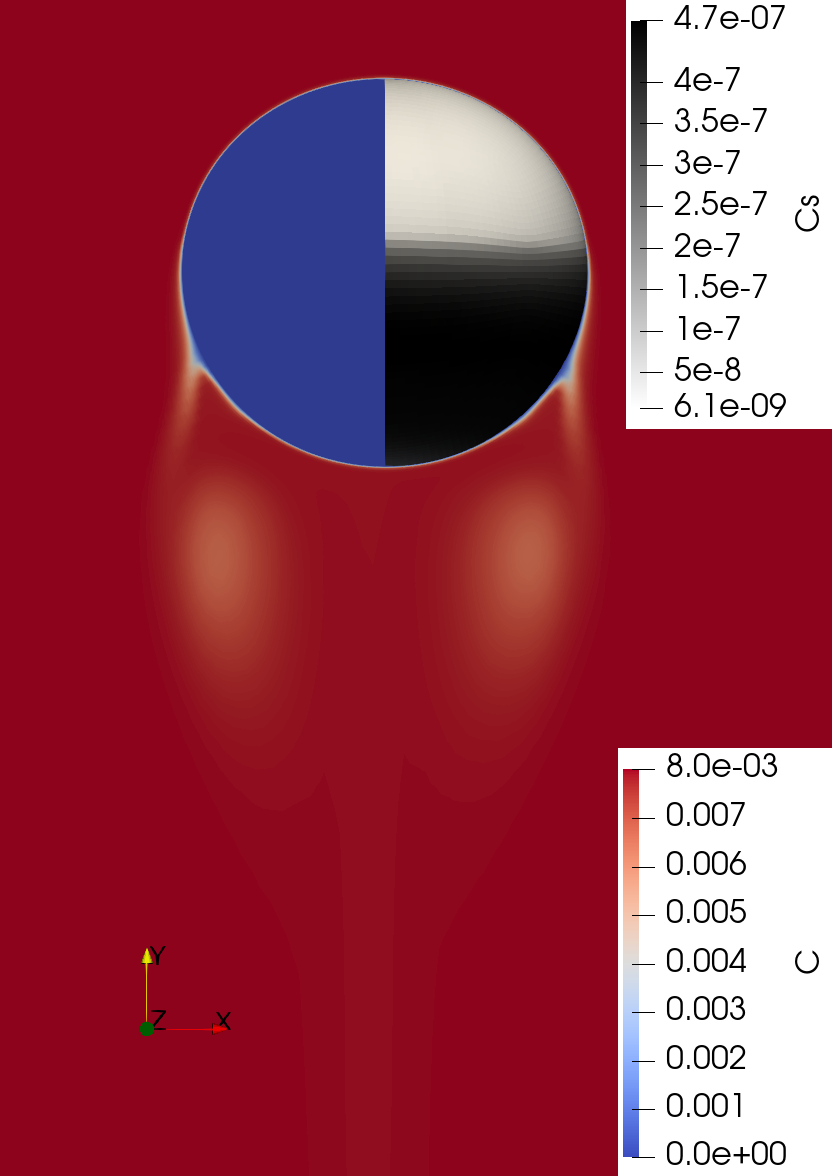}
        \caption{$c_{\infty} = $ \SI{8e-3}{mol/m^3}}
        % \label{fig:rotating_ParamStudy_RoC}
    \end{subfigure}
    \begin{subfigure}{.3\textwidth}
        \centering
        \includegraphics[width=0.95\linewidth]{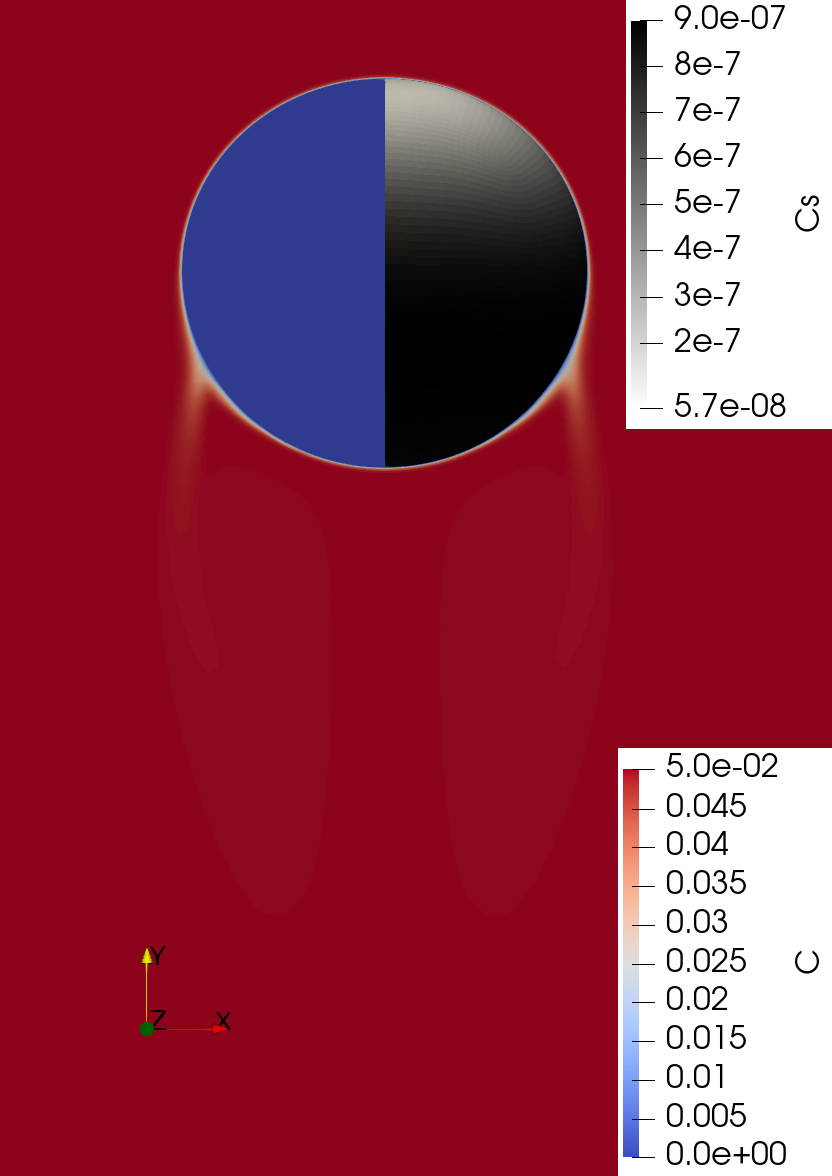}
        \caption{$c_{\infty} = $ \SI{5e-2}{mol/m^3}}
        % \label{fig:rotating_ParamStudy_RoC}
    \end{subfigure}
    \caption{Surfactants on the surface and in the bulk phases at t = \SI{0.1}{s} for different concentration levels}
    \label{fig:risBubble_surfactDistr}
\end{figure}

\Cref{fig:risBubble_surfactDistr} shows the surfactant distribution in the bulk phases and on the surface after \SI{0.1}{s} for the 3~different cases with surfactant.
While a thin boundary layer can be observed in all cases, the area where it detaches and the amount accumulated on the interface vary greatly.
In all cases, the surfactant accumulates at the rear part, where it immobilizes the surface, as can be concluded from the low velocities there in \cref{fig:risBubble_velos}.
This changes the flow field significantly.
In the clean case, there is no visible detachment of the bubble wake, whereas in the case with the highest concentration, there is detachment and a recirculation zone bigger than the bubble itself.
Another major difference between the cases is the shape of the bubble: the clean bubble deforms into a spheroidal form, % Rotationsellipsoid
whereas in the cases with high concentrations the bubble remains almost spherical.

\begin{figure}[!htbp]
    \centering
    \begin{subfigure}{.24\textwidth}
        \centering
        \includegraphics[width=0.95\linewidth]{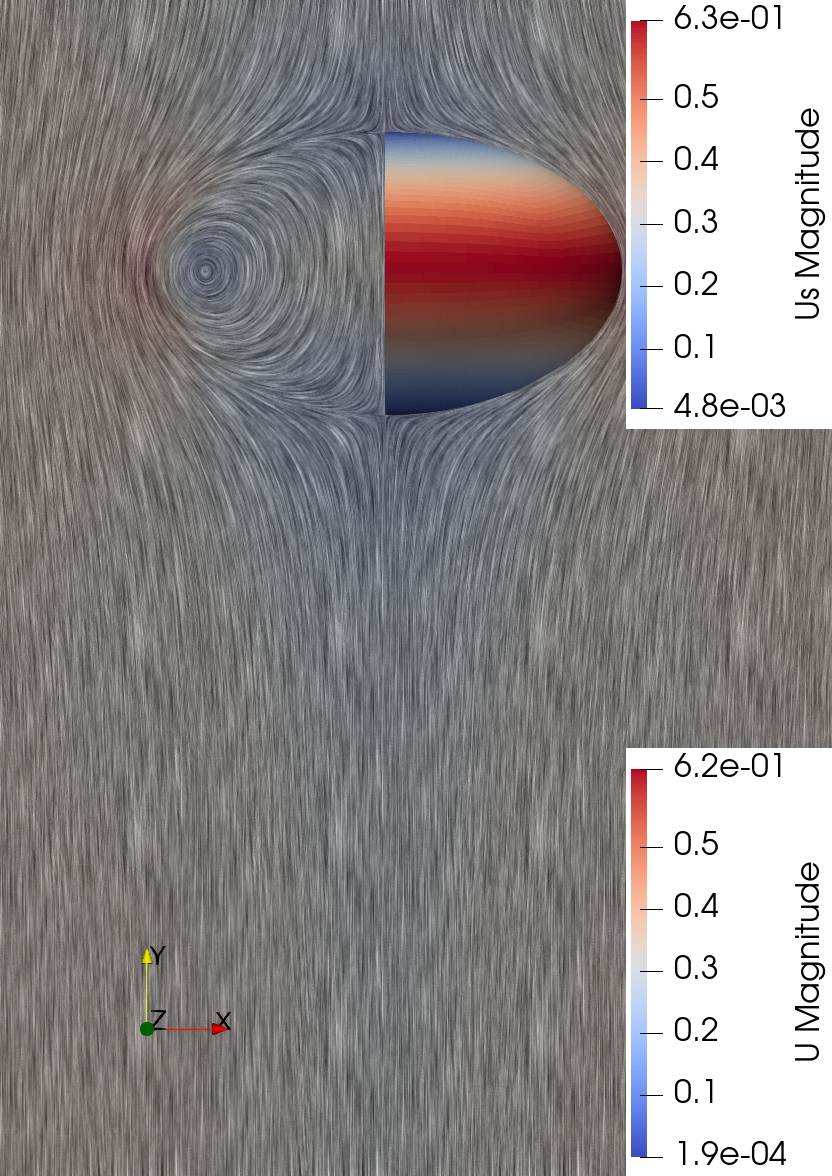}
        \caption{$c_{\infty} = $ \SI{0}{mol/m^3}}
        % \label{fig:rotating_ParamStudy_errorNorms}
    \end{subfigure}
    \begin{subfigure}{.24\textwidth}
        \centering
        \includegraphics[width=0.95\linewidth]{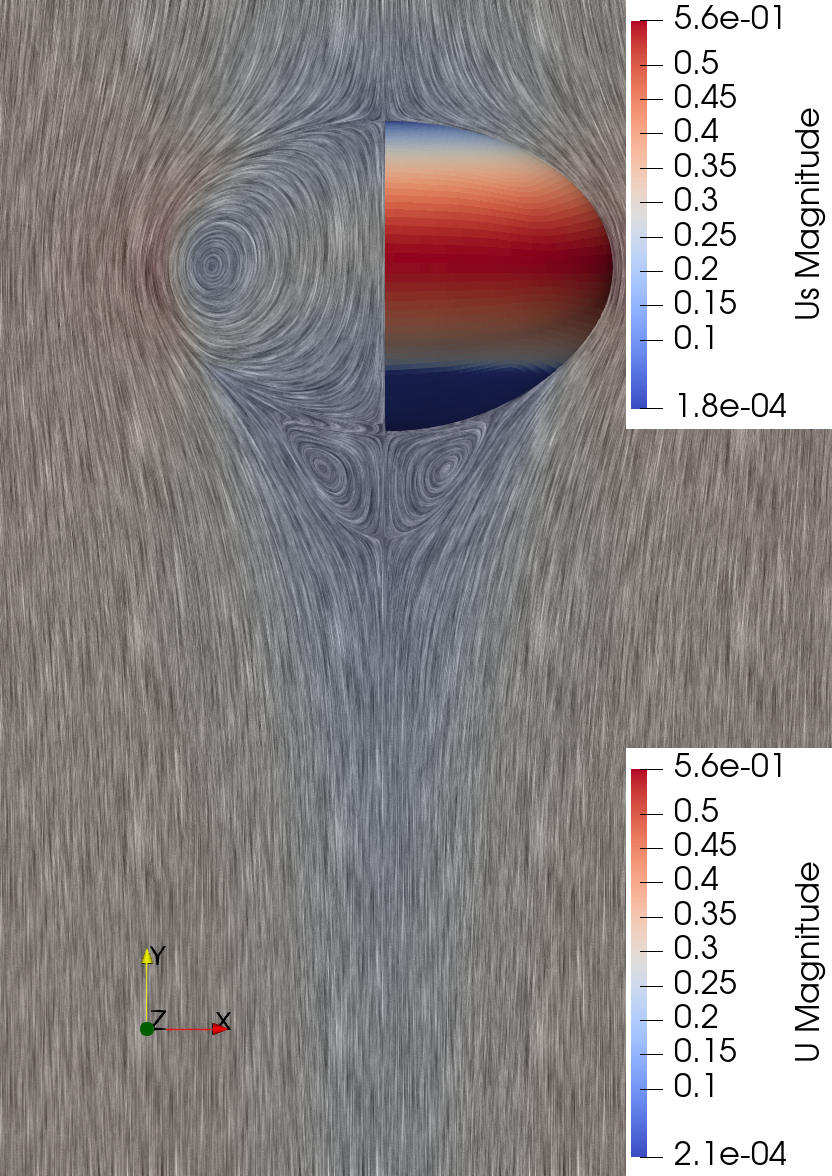}
        \caption{$c_{\infty} = $ \SI{2e-3}{mol/m^3}}
        % \label{fig:rotating_ParamStudy_errorNorms}
    \end{subfigure}
    \begin{subfigure}{.24\textwidth}
        \centering
        \includegraphics[width=0.95\linewidth]{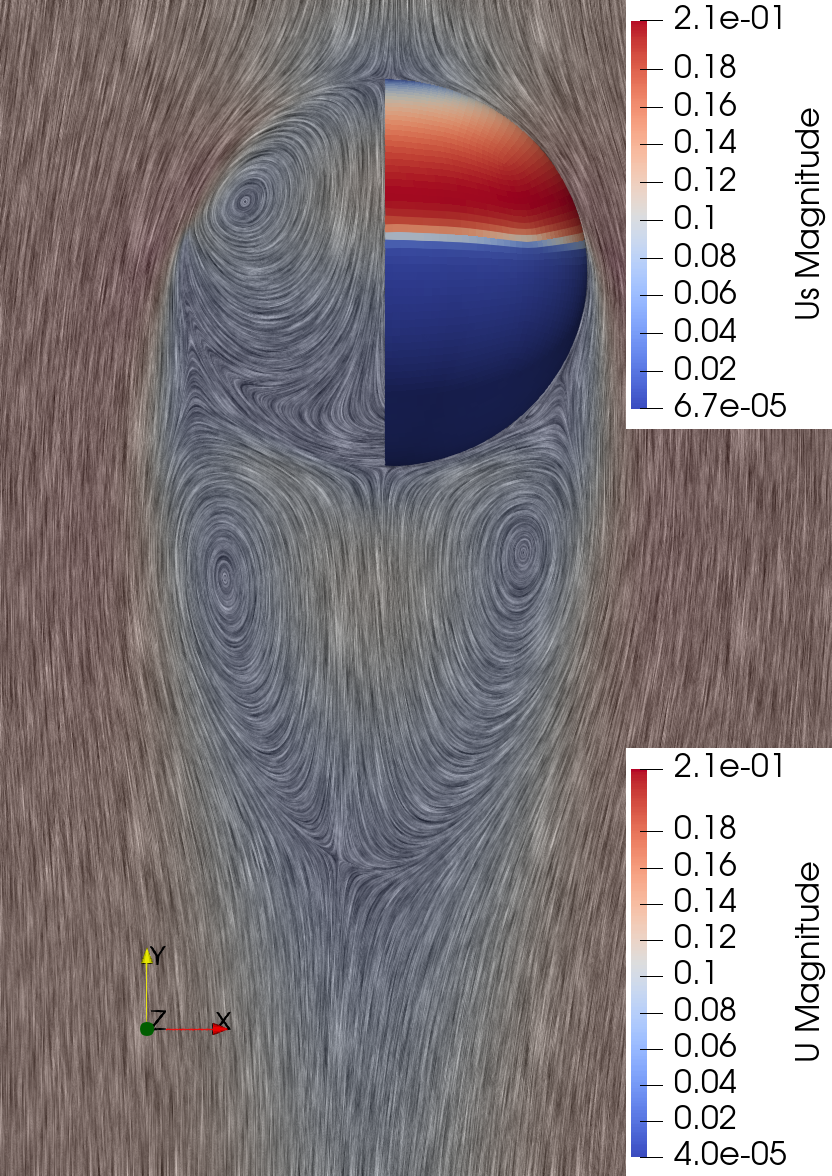}
        \caption{$c_{\infty} = $ \SI{8e-3}{mol/m^3}}
        % \label{fig:rotating_ParamStudy_RoC}
    \end{subfigure}
    \begin{subfigure}{.24\textwidth}
        \centering
        \includegraphics[width=0.95\linewidth]{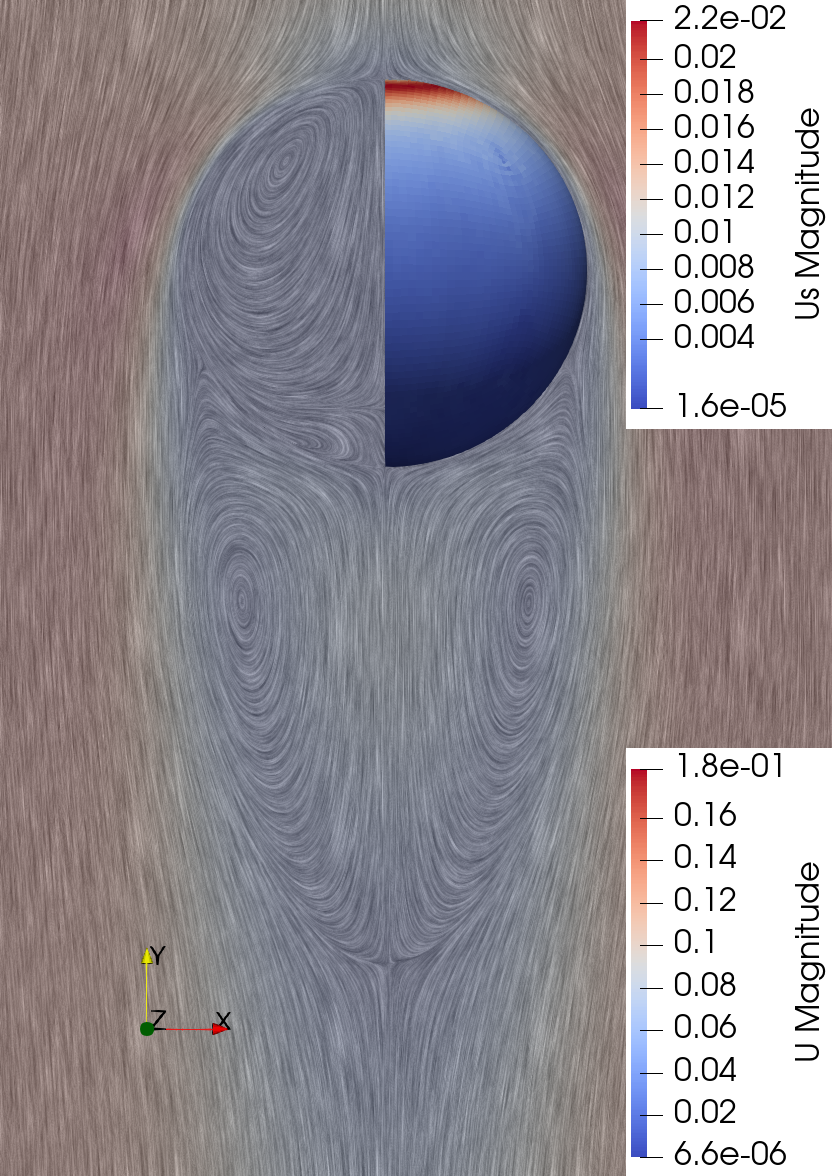}
        \caption{$c_{\infty} = $ \SI{5e-2}{mol/m^3}}
        % \label{fig:rotating_ParamStudy_RoC}
    \end{subfigure}
    \caption{Velocities on the surface and in the bulk phases at t = \SI{0.1}{s} for different concentration levels}
    \label{fig:risBubble_veloFields}
\end{figure}

We have shown for a rising bubble in \cref{fig:risBubble_velos,fig:risBubble_surfactDistr,fig:risBubble_veloFields}, that our framework can effectively capture the expected results in the rising velocity of the bubble, the surfactant concentration in the bulk and on the interface, as well as their influence on the bubble shape.

\section{Conclusions}
\label{sec:conclusions}

We provide a thoroughly automatically tested implementation of the unstructured Finite-Volume Arbitrary Lagrangian / Eulerian (ALE) Interface-Tracking (IT) method for simulating incompressible, immiscible two-phase flows \otherChange{with surfactants and Subgrid-Scale (SGS) modeling of interfacial mass transfer}
as an OpenFOAM module.

\otherChange{We provide a modular and easily extendible SGS} model hierarchy within an \otherChange{ALE-IT} OpenFOAM module, with configurable output parameters for in-depth analysis. The SGS model's ability to accurately capture thin boundary layers and steep gradients was demonstrated, alongside the critical need for flux correction not only at the interfaces but also across adjacent cell boundaries. Furthermore, the Interface Tracking module's proficiency in accurately simulating surfactants on moving fluid interfaces interfaces and their impact on surface tension and Marangoni flow was demonstrated.
\ReviewerTwo{We demonstrate the capabilities of our ALE-IT implementation for simulating complex hydrodynamics with mass transfer by simulating a rising bubble with soluble surfactants.
%\cref{fig:risBubble_velos,fig:risBubble_surfactDistr,fig:risBubble_veloFields}, 
Our framework is able to reproduce the expected results, such as the rising velocity of the bubble, the surfactant concentration in the bulk and on the interface, as well as the surfactants influence on the bubble shape.}

\ReviewerTwo{Ongoing work includes ensuring consistent contact line kinematics \citep{fricke_kinematics_2018,Fricke2019} for the ALE-IT framework to improve simulation accuracy for contact line kinematics and dynamics, and extensions towards tetrahedral meshes for stronger mesh deformation and topological changes \citep{menon_parallel_2015}.
}

\section{Acknowledgments}

Moritz Schwarzmeier and Tomislav Mari\'{c} would like to thank the Federal Government and the Heads of Government of the Länder, as well as the Joint Science Conference (GWK), for their funding and support within the framework of the NFDI4Ing consortium. Funded by the German Research Foundation (DFG) - project number 442146713.

\medskip

Funded by the German Research Foundation (DFG)
– Project-ID 265191195 – SFB 1194.

\medskip

Part of this work was funded by the Hessian Ministry of Higher Education, Research, Science and the Arts - cluster project Clean Circles.

\medskip

\otherChange{
The authors gratefully acknowledge the computing time provided to them on the high-performance computer Lichtenberg~II at TU~Darmstadt, funded by the German Federal Ministry of Education and Research (BMBF) and the State of Hesse.}

\appendix
\label{sec:appendix}

\section{Automatic testing}
\label{secSub:AutoTesting}

\begin{figure}[!htbp]
    \centering
    \includegraphics[width=\textwidth]{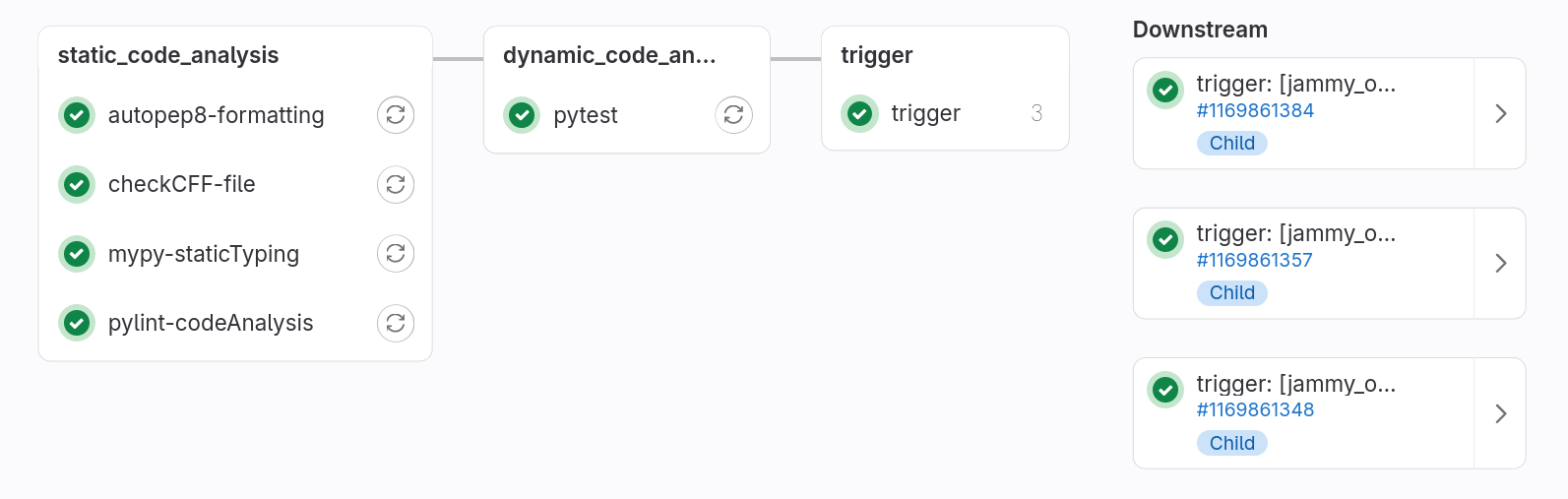}
    \caption{Upstream CI pipeline}
    \label{fig:CI_pipeline_upstream}
\end{figure}

\begin{figure}[!htbp]
    \centering
    \includegraphics[width=\textwidth]{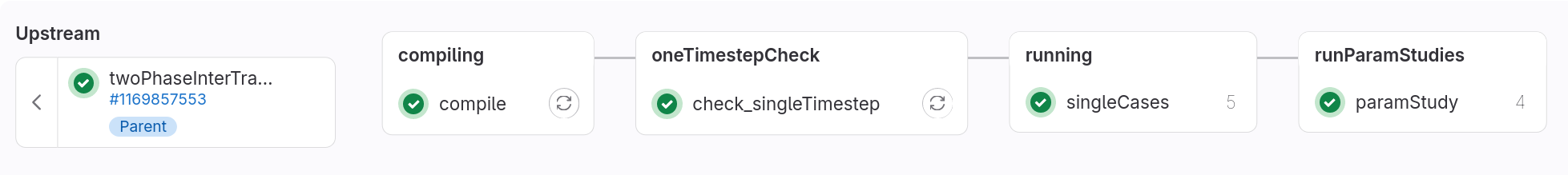}
    \caption{Downstream CI pipeline}
    \label{fig:CI_pipeline_downstream}
\end{figure}

We have set up our test cases in a Continuous Integration (CI) pipeline. The pipeline is depicted in \cref{fig:CI_pipeline_upstream,fig:CI_pipeline_downstream}.
In our approach we follow the practices presented in \cite{maric_workflow_2022,maric_pragmatic_2023}.
This means the tests are run and evaluated automatically in the CI with Jupyter Notebooks. They also produce the results and graphs included in this work.
A job in a succeeding stage to the downstream pipeline has been omitted from \cref{fig:CI_pipeline_downstream} for the sake of readability.
Said job collects and bundles all those artifacts from the two previous stages.
%the simulations in the stages \parameter{running} and \parameter{runParamStudies}.

In \cite{maric_workflow_2022,maric_pragmatic_2023} CI is focused on tests and their automatic evaluation. We have added tests to the CI pipeline, that check certain additional aspects of the repository, like e.g.\ formatting and the formal correctness of the \file{CITATION.cff}-file. We also build an image within the CI with OpenFOAM and other needed packages. This job can not be seen in \cref{fig:CI_pipeline_upstream}, since it is only run once a new OpenFAOM version is available. It is located before the trigger job for the downstream pipelines.

We further expanded the CI to be able to test the code for multiple OpenFOAM versions, which at the moment is being done for the latest three versions OpenFOAM \parameter{v2312}, \parameter{v2306} and~\parameter{v2212}. In feature-branches only the latest version is tested to save resources.

Another feature of the CI pipeline is the addition of a test, that checks the OpenFOAM log-files for warnings and errors, where tutorials added to the repository are automatically detected and run for a single timestep. This almost ensures, that there are no errors in the set-up of test cases and would also uncover some of errors in the code, that can only be detected, when OpenFOAM is run. The automatisation of this test in finding and running all simulation cases is especially valuable. Since running all of the simulations for only one timestep needs a very limited amount of time it can be checked, that changes in the code don't corrupt any case before commiting and pushing to the central repository.
The automated finding of cases also prevents forgetting to add a simulation case to the CI (to this testing method) and the automated execution and error finding gently forces all participants of the repository to set up their cases to run flawlessly and fully automatised, e.g.\ they need to include everything needed to run a case in the \file{Allrun} script. This ensures all the cases provided with the code can be run without debugging. % debugging them first / fixing them

Another topic not covered in the previously mentioned references is that for repetitive tasks, like reading dictionary entries, reading files or calculating error norms needed for the evaluation a set of python functions is provided. This set of functions is also tested automatically with the testing framework pytest as part of the CI pipeline and precedes the more computationally expensive OpenFOAM-specific jobs, as can be seen in \cref{fig:CI_pipeline_upstream}.

% Bibliography
%\section*{References} % Uncomment on Overleaf. 
\bibliography{bibliography}

\end{document}